\documentclass[prd,10pt,notitlepage,nofootinbib,showpacs,showkeys]{revtex4-1}
\usepackage{latexsym,graphicx,amssymb,amsmath,mathrsfs}
\usepackage{setspace,bm}

\newcommand{\be}{\begin{equation}}
\newcommand{\ee}{\end{equation}}  
\newcommand{\bea}{\begin{eqnarray}}
\newcommand{\eea}{\end{eqnarray}}  
\renewcommand{\k}{{\bf k}}
  
\newcommand{\p}{{\bf p}}
\newcommand{\vac}{\textrm{(v)}}

\newcommand{\Gv}{G^{\,\textrm{(v)}}}
\newcommand{\Gm}{G^{\,\textrm{(m)}}}
\newcommand{\Gr}{G^{\,\textrm{(r)}}}
\newcommand{\Dv}{\Delta^{\textrm{(v)}}}
\newcommand{\Dm}{\Delta^{\textrm{(m)}}}
\newcommand{\Dr}{\Delta^{\textrm{(r)}}}
\newcommand{\Sigmar}{\Sigma^{\textrm{(r)}}}
\newcommand{\SigmaO}{\Sigma^{\textrm{(0)}}}
\newcommand{\Sigmav}{\Sigma^{\textrm{(v)}}}

\newcommand{\rt}[1]{{}}    

\begin{document}
\allowdisplaybreaks

\author{Gergely Fej\H{o}s}
\email{geg@ludens.elte.hu}
\affiliation{Department of Atomic Physics, E{\"o}tv{\"o}s University, H-1117 Budapest, Hungary}

\author{Zsolt Sz{\'e}p}
\email{szepzs@achilles.elte.hu}
\affiliation{Centre de Physique Th{\'e}orique, Ecole Polytechnique, CNRS, 91128 Palaiseau Cedex, France.}
\altaffiliation{On leave from Statistical and Biological Physics Research Group
of the Hungarian Academy of Sciences, H-1117 Budapest, Hungary.}

\title{ Broken symmetry phase solution of the $\phi^4$ model at
   two-loop level of the $\Phi$-derivable approximation}

\begin{abstract}
The set of coupled equations for the self-consistent propagator and
the field expectation value is solved numerically with high accuracy
in Euclidean space at zero temperature and in the broken symmetry
phase of the $\phi^4$ model.  Explicitly finite equations are derived
with the adaptation of the renormalization method of van Hees and
Knoll \cite{vanHees:2001ik} to the case of nonvanishing field
expectation value. The set of renormalization conditions used in this
method leads to the same set of counterterms obtained recently by
Patk{\'o}s and Sz{\'e}p in \cite{patkos08}. This makes possible the direct
comparison of the accurate solution of explicitly finite equations
with the solution of renormalized equations containing counterterms.
The numerically efficient way of solving iteratively these latter
equations is obtained by deriving at each order of the iteration new
counterterms which evolve during the iteration process towards the
counterterms determined based on the asymptotic behavior of the
converged propagator.  As shown at different values of the coupling,
the use of these evolving counterterms accelerates the convergence of
the solution of the equations.
\end{abstract}

\pacs{02.60.Cb, 11.10.Gh, 12.38.Cy}
\keywords{Renormalization; 2PI formalism; Solution of integral equations}

\maketitle

\section{Introduction and motivation}

The $\Phi$-derivable approximation \cite{ward60,Cornwall:vz} (also
called the two-particle-irreducible (2PI) approximation) represents a
self-consistent and systematically improvable approximation scheme,
successfully applied in both out-of-equilibrium and equilibrium
contexts.  In the first case, it overcomes the secularity problem
encountered with methods not using a self-consistent propagator (see
the references given in \cite{cooper04}) and makes possible the
description of late time dynamics of scalar
\cite{Berges:2000ur,Aarts:2001qa,arrizabalaga05,Borsanyi:2008ar} and
fermionic \cite{Berges:2002wr} quantum fields.  In the second case, it
represents an adequate framework for the study of the phase transition
in quantum field theories, where some systematic partial resummation
of the perturbative series has to be implemented. This is needed
because the perturbative expansion around the free theory cannot
capture the development of collective phenomena and the change in the
physical spectrum occurring at high temperature, where the temperature
could compensate the smallness of the coupling \cite{dolan74}.  More
sophisticated resummations can be obtained by combining the 2PI
formalism with an expansion in number of flavors
\cite{Berges:2001fi,Aarts:2002dj}.

In the past few years much attention was devoted to understand the
renormalization in the 2PI formalism.  In \cite{vanHees:2001ik} the
temperature dependent subdivergences of the self-consistent propagator
equation were localized in the symmetric phase of the $\phi^4$ model
by expanding the propagator and the self-energy around the
corresponding zero temperature quantities. It was shown that the
divergences are resummed by a Bethe-Salpeter-type equation and that
the self-energy can be made finite if one renormalizes both a
four-point function satisfying the Bethe-Salpeter equation and its
kernel. Imposing renormalization conditions, the renormalization of
the self-consistent propagator equation at ${\cal O}(\lambda^2)$
truncation of the 2PI effective action of the above model was
performed in the symmetric phase in \cite{VanHees:2001pf}. In
\cite{vanHees:2001ik} no attention was paid to subdivergences
appearing at zero temperature, although as shown in
\cite{Blaizot:2003br, Blaizot:2003an}, the same structure of
subdivergences pointed out in \cite{vanHees:2001ik} appears already
there. The use of a finite number of counterterms in
\cite{Blaizot:2003br, Blaizot:2003an} made transparent that once the
theory is renormalized at $T=0$ it remains finite at finite
temperature, as expected. The extension of the renormalization
method of \cite{vanHees:2001ik} to theories with
spontaneously broken symmetries was treated in
\cite{vanHees:2002bv}, where it was explicitly applied to the $O(N)$
model at Hartree level truncation of the 2PI effective action. In
\cite{Berges:2005hc} it was shown for a generic truncation of the 2PI
effective potential that the renormalization at nonvanishing field
expectation value can be achieved by renormalizing the theory in the
symmetric phase.  The quantities which need to be renormalized in the
symmetric phase to render the propagator and field equations finite
are the various two- and four-point functions and the 2PI kernels of
the Bethe-Salpeter equations which are satisfied by the former
quantities and which resum different types of subdivergences. These
quantities arise naturally by taking sequential derivatives of the
so-called 2PI-resummed effective action with respect to the field. The
removal of the divergences by appropriate counterterms is achieved by
imposing renormalization and consistency conditions \footnote{This
  distinction was introduced in \cite{Reinosa:2009tc} to differentiate
  the condition imposed one the 2PI vertex function obtained as
  field-derivatives of the inverse propagator which extremizes the 2PI
  effective action through the stationarity condition, from the
  condition which is imposed on the 2PI-resummed vertex function
  defined as field derivatives of the 2PI-resummed (1PI) effective
  action, and which differs from the 2PI vertex function only due to
  the approximation made on the 2PI effective action.} on the 2PI
kernels and the two- and four-point functions. This diagrammatic
renormalization procedure was extended to models containing fermionic
\cite{Reinosa:2005pj} and gauge degrees of freedom
\cite{Reinosa:2006cm}, and the applicability of the 2PI
renormalization method in the study of the time evolution of
out-of-equilibrium scalar fields was shown in \cite{Borsanyi:2008ar}.
Working in the broken symmetry phase of one- and multicomponent
scalar models, an alternative method for obtaining the counterterms up
to skeleton order ${\cal O}(\lambda^2)$ truncation of the 2PI
effective action was given in \cite{Fejos:2007ec,patkos08}.  The
renormalization of the $O(N)$ model in the $1/N$ expansion of the 2PI
effective action was achieved in
\cite{Berges:2005hc,cooper04,Fejos:2009dm} using the original fields
and also the auxiliary field method.  A renormalization method was
developed in \cite{Jakovac:2006dj,Jakovac:2006gi} which, instead of
considering the divergences of the usual Feynman diagrams and the
Bethe-Salpeter equations resumming them, uses a specific resummation
scheme at a given temperature where the model is parametrized. At a
different temperature the method uses matching conditions at
asymptotic momenta.

In an equilibrium setting and beyond the Hartree approximation there
are relatively few papers reporting on numerical solutions of
renormalized 2PI equations, even in scalar models.  In some
phenomenological studies \cite{Roder:2005vt,roder05} the
renormalization is not done, {\it e.g.} in \cite{Roder:2005vt} the
$O(N)$ model is solved in the Minkowski space with some approximations
by taking into account momentum-dependent corrections only in the
imaginary part of the self-energy and treating the tadpole with a UV
cutoff.

The application of the 2PI method did not went yet beyond the
demonstration of its features in the simplest models (e.g. the
$N$-component real scalar model with $O(N)$ symmetry, in many cases
restricted to $N=1$).  The explicitly finite self-consistent
propagator equation of the $\phi^4$ model was solved including the
setting-sun diagram at vanishing field expectation value and at finite
temperature in \cite{VanHees:2001pf}.  The renormalized $O(N)$ model
was solved at zero temperature and at vanishing field expectation
value within the bare-vertex approximation of the auxiliary field
formalism in \cite{cooper04}.  The pressure of the one-component
$\phi^4$ model was calculated in Euclidean space and finite
temperature in \cite{Berges:2004hn} solving a renormalized 3-loop
effective action.  Solving the renormalized field and propagator
equation of this model in Minkowski space the phase transition was
studied in \cite{arrizabalaga06}. It was found that including the
field-dependent two-loop skeleton diagram at the level of the 2PI
effective potential results in a second order phase transition
\cite{arrizabalaga06}.\footnote{It was proven analytically in
  \cite{Reinosa:2011ut} that in Hartree approximation the phase
  transition cannot be of second order.}  The study of thermal
properties of the spectral function of the renormalized $\phi^4$
theory in the symmetric phase was reported in \cite{Jakovac:2006gi}.

However, especially in the above listed finite temperature studies,
little technical details are given on the numerics, in particular, the
rate of convergence of the iterative steps leading to self-consistent
solutions.  This fact does not allow us to easily infer the accuracy
of the methods used in obtaining the results. In our opinion, the lack
of standardized, well-tested algorithms might be the main reason
\rt{of disconverging} for the rather scarce number of applications
  of the 2PI-approximation in the study of the thermodynamics of
relevant field theoretical models.  Our aim in the present work is to
quantify the accuracy of iterative solutions and to compare different
algorithms. We also would like to make connections between different
renormalization methods.  Our highly accurate solutions could serve as
benchmark for more powerful methods to be used for a finite
temperature solution of these equations.

The outline of the paper is as follows. In Sec.~II we introduce the
model and derive the coupled set of self-consistent gap and field
equations.  In Sec.~III we discuss their renormalization by adapting
to broken symmetry phase the renormalization method of
\cite{vanHees:2001ik,vanHees:2002bv}.  We isolate the quantities which
need to be renormalized and imposing renormalization conditions we
recover, on the one hand, the set of counterterms derived with a
different method in \cite{patkos08} and, on the other hand, derive
explicitly finite propagator and field equations.  In Sec.~IV we give
the algorithms for solving at zero temperature and in Euclidean space
the set of finite equations and the equations containing explicitly
the counterterms.  We demonstrate that their numerical solution
coincide and investigate the efficiency of the two methods.  We show
that in the case of solving the equations containing the counterterms
faster convergence of the iterative algorithm is obtained if the
counterterms are rederived in order to allow them to evolve during the
iterative procedure.  Sec.~V is devoted to conclusions.

\section{The 2PI effective potential at two-loop field-dependent level}

\begin{figure}[t]
\begin{center}
\includegraphics[keepaspectratio,width=0.6\textwidth,angle=0]{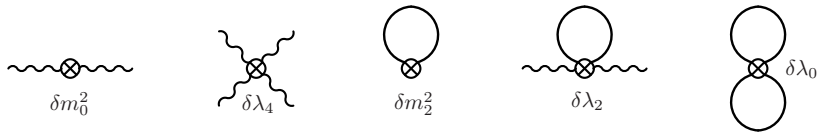}
\caption{The counterterm diagrams corresponding to the terms of 
$V_{\rm ct}[v,G]$ defined in (\ref{Eq:l2-2PI-Vct}). A wiggly line represents $v$, while a plain one corresponds to the full propagator $G$. Symmetry factors are not indicated.}
\label{Fig:cts}
\end{center}
\end{figure}

The 2PI effective potential of the real one-component $\phi^4$ model
at the field-dependent two-loop truncation level can be given in the
broken symmetry phase, characterized by the vacuum expectation value
$v,$ in the following form 
\cite{Berges:2005hc,arrizabalaga05,arrizabalaga06,patkos08}:
\bea
V[v,G]&=&\frac{1}{2}m^2v^2+\frac{\lambda}{24}v^4
-\frac{i}{2}\int_P\big[\ln G^{-1}(P)+D^{-1}(P)G(P)\big]
+\frac{\lambda}{8}\left(\int_P G(P)\right)^2
\nonumber\\
&&-\frac{i\lambda^2}{12}v^2\int_K\int_P G(K)G(P)G(K-P)
+V_\textrm{ct}[v,G].
\label{Eq:l2-2PI_V}
\eea 
The expression above is written in Minkowski space following the
conventions of Ref. \cite{Peskin-book}: the tree-level propagator is
$D(P)=i/(P^2-m^2-\lambda v^2/2)$ with $P=(p_0,\p)$ denoting the
four-momentum and the metric is such that $P^2=p_0^2-\p^2.$ The
counterterm functional is given by
\footnote{Unfortunately in Ref.~\cite{patkos08} $\delta m_0^2$ and 
$\delta m_2^2$ were interchanged compared to the notation originally introduced 
in Ref.~\cite{Berges:2005hc}. In this work we will stick to the notation of 
Ref.~\cite{patkos08}.}
\be
V_\textrm{ct}[v,G]=\frac{1}{2}\delta m^2_0 v^2+\frac{\delta\lambda_4}{24}v^4+
\frac{1}{2}\left(\delta m_2^2
+\frac{\delta\lambda_2}{2}v^2\right)\int_P G(P)
+\frac{\delta\lambda_0}{8}\left(\int_P G(P)\right)^2,
\label{Eq:l2-2PI-Vct}
\ee 
where the different terms are represented graphically in
  Fig.~\ref{Fig:cts}.  As discussed in \cite{Berges:2005hc}, the
origin of two different mass counterterms and three coupling
counterterms stems from the fact that within the 2PI formalism it is
possible to define two two-point functions and three four-point
functions, which do not necessarily coincide within a given truncation
of the 2PI effective potential. In writing (\ref{Eq:l2-2PI_V}) it is
implicitly assumed that, despite the possible different bare masses
and counterterms their finite part agree, that is there is only one
renormalized mass parameter $m^2$, which in the symmetry breaking
phase is usually negative, and only one renormalized coupling
parameter~$\lambda$.

The stationarity conditions $\delta V[v,G]/\delta G=0$ and  
$\delta V[v,G]/\delta v=0$ give a self-consistent equation for the 
full two-point function and the equation for the vacuum expectation
value, {\it i.e.} the field equation: 
\bea
\label{Eq:full_prop}
i G^{-1}(P)&=&P^2-\Sigma(P),\\
\label{Eq:EoS}
0&=&v\left(m^2+\delta m_0^2+\frac{1}{6}(\lambda+\delta\lambda_4)v^2
+\frac{1}{2}(\lambda+\delta\lambda_2)T[G]+\frac{\lambda^2}{6}S(0,G)\right),
\eea
where the self-energy $\Sigma(P)$ is given by
\be
\Sigma(P)=m^2+\delta m_2^2+\frac{1}{2}(\lambda+\delta\lambda_2)v^2
+\frac{1}{2}(\lambda+\delta\lambda_0) T[G]
+\frac{1}{2}\lambda^2 v^2 I(P,G).
\label{Eq:Sigma_def}
\ee
Note that we call $\Sigma(P)$ the self-energy, 
although its usual definition used in the 2PI
formalism does not contain the mass parameter and the corresponding
mass counterterm. The tadpole $T[G]$, bubble $I(P,G)$ 
and setting-sun at vanishing external momentum $S(0,G)$
which appear above are defined as follows:
\begin{subequations}
\bea
T[G]&=&\int_K G(K),
\label{Eq:T_def}
\\
I(P,G)&=&-i \int_K G(K)G(P-K),
\label{Eq:I_def}
\\
S(0,G)&=&-i \int_K\int_Q G(K) G(Q) G(-K-Q).
\label{Eq:SS_def}
\eea
\label{Eq:TBSS}
\end{subequations}
\vspace{-2mm}   

\noindent
In \cite{patkos08} the divergences of these integrals were calculated
by expanding the full propagator around the auxiliary propagator
\be
G_0(P)=\frac{i}{P^2-M_0^2}, 
\label{Eq:G0}
\ee
with the mass scale $M_0$ playing the role of the renormalization
scale. The counterterms absorbing these divergences were obtained in
\cite{patkos08} by requiring the separate cancellation of the
divergent coefficients of $v^0$, $v^2$ and of the environment ($T$ and
$v$) dependent finite function $T_F[G]$, representing the finite part
of the tadpole integral, both in the propagator and field
equations. They are given in Appendix~\ref{app:cutoff} in terms of
some explicitly calculated integrals.

\section{Derivation of explicitly finite equations \label{sec:finite_equation}}

In this section we adapt the renormalization method used at finite
temperature in \cite{vanHees:2001ik} to the broken symmetry phase at
zero temperature. We mention that the renormalization of scalar models
displaying spontaneously broken symmetries was discussed generally in
the context of the 2PI approximations in
\cite{vanHees:2002bv}. Here, we separate the field dependence from the
divergent quantities in the same way as the temperature dependence was
separated in \cite{vanHees:2001ik} and give the renormalization
conditions which, when imposed on the divergent quantities, lead on
the one hand to exactly the counterterms determined in \cite{patkos08}
and summarized in (\ref{Eq:all_counterterms}), and on the other hand
to explicitly finite equations, with no reference to any counterterms
at all.  We note that our counterterms have exactly the same structure
as those used in \cite{arrizabalaga06} and which were derived using
the renormalization method of
\cite{Berges:2005hc}. \footnote{Unfortunately the counterterms used in
  \cite{arrizabalaga06} appeared only on the corresponding poster.}
In the next section we will check explicitly that the solution of the
propagator and field equation obtained using counterterms agrees with
the solution of the explicitly finite equations to be derived below.

\subsection{Renormalization of the propagator equation}

We begin by splitting the propagator into ``vacuum'' and ``matter'' parts,
\be
G(P)=\Gv(P)+\Gm(P),
\label{Eq:G_split}
\ee
where the vacuum part refers to a propagator defined in the symmetric
phase ($v=0$), while the matter propagator includes all explicit and
implicit dependence on $v.$ The truncation of the 2PI effective
potential studied in this work leads to a momentum independent
self-energy for $v=0$ which means that the vacuum propagator can be
simply parametrized in terms of an effective positive mass $M_0$, as
in (\ref{Eq:G0}), so that $\Gv(P)\equiv G_0(P)$. Note that we need
this propagator with positive mass squared $M_0^2$ because in the
broken symmetry phase $m^2,$ the renormalized mass parameter of the
Lagrangian could be negative.

As a consequence of the splitting (\ref{Eq:G_split}), 
the self-energy (\ref{Eq:Sigma_def}) is decomposed into three parts:
\bea
\Sigma(P)=\Sigmav+\SigmaO(P)+\Sigmar(P),
\label{Eq:sigma_decomp}
\eea
where the explicit expression of the different parts are given below.
The vacuum part is independent of the momentum, depends only on the 
vacuum propagator and has divergence degree 2. The last two pieces of 
the self-energy appearing in (\ref{Eq:sigma_decomp}) are $v$ dependent 
and are defined as follows. The part which does not result in any 
divergence when the full propagator is expanded around $\Gv$ is called 
regular part and is denoted by $\Sigmar$, while the remaining part, 
having divergence degree 0, is denoted by $\SigmaO$. The splitting of
the field-dependent part is somewhat arbitrary. Using (\ref{Eq:G_split})
in (\ref{Eq:Sigma_def}), the different parts of
(\ref{Eq:sigma_decomp}) are identified as
\begin{subequations}
\label{Eq:sigmas}
\bea
\label{Eq:sigma-vac}
\Sigmav&=&M_0^2+\delta m_{2,A}^2
+\frac{\lambda+\delta \lambda_0}{2}\int_K \Gv(K),  \\
\label{Eq:sigma-0}
\SigmaO(P)&=&m^2-M_0^2+\delta m_{2,B}^2
+\frac{v^2}{2}\Lambda_2^{\vac}(0,P)
+\frac{\lambda+\delta \lambda_0}{2}\int_K \Gm(K),\ \  \\
\label{Eq:sigma-r}
\Sigmar(P)&=&-i\lambda^2 v^2\int_K \Gm(K) \Gv(K+P)
-i\frac{\lambda^2 v^2}{2}\int_K \Gm(K)\Gm(K+P),
\eea
\end{subequations}
where $\Lambda_2^{\vac}(0,P)$ is defined as
\be
\Lambda_2^{\vac}(0,P)=\lambda+\delta \lambda_2
-i\lambda^2\int_K \Gv(K)\Gv(K+P).
\label{Eq:Lambda2_vac}
\ee
We will see that with the above choice for the different pieces of the
self-energy the expression of the counterterms
(\ref{Eq:all_counterterms}) can be obtained through simple
renormalization conditions. The mass counterterm $\delta m^2_2$ was
split into two parts: $\delta m^2_{2,A}$ is responsible for the
renormalization of the vacuum part, while $\delta m^2_{2,B}$ has to
remove the divergence generated by the $\SigmaO$ dependence of
$\Gm$. In order to see how this latter divergence proportional to
$\SigmaO$ emerges, we expand the propagator around the vacuum part:
\bea
G(P)&=&
\frac{i}{P^2-\Sigmav-\SigmaO(P)-\Sigmar(P)}=
\frac{i}{P^2-\Sigmav(P)}\left(1+\frac{\SigmaO(P)+
\Sigmar(P)}{P^2-\Sigmav}+\dots\right) \nonumber\\
&=&\Gv(P)-i\Big(\Gv(P)\Big)^2\SigmaO(P)+\Gr(P).
\label{Eq:prop}
\eea
The term proportional to $\SigmaO$ is ${\cal{O}}(1/P^4)$ (up to logs),
therefore it gives divergent contribution after integrating over the
momentum, while the regular part of the propagator goes with
${\cal{O}}(1/P^6)$ and leads to a finite contribution upon integration
over the momentum.  The sum of these two terms is the previously
introduced matter part
\be
\Gm(P)=-i\Big(\Gv(P)\Big)^2\SigmaO(P)+\Gr(P).
\label{Eq:G-mat}
\ee
As already announced at the beginning of this subsection, 
the first renormalization condition is 
\be
\Sigmav=M_0^2,
\label{Eq:ren_cond_sigma-vac}
\ee 
which determines $\delta m_{2,A}^2$ through the relation
\be
\delta m_{2,A}^2=-\frac{\lambda+\delta \lambda_0}{2}T_d^{(2)},
\label{Eq:delta-m22A}
\ee
where $T_d^{(2)}$ is defined in (\ref{Eq:Td2_Td0_S0}). 

Now we turn to the  renormalization of $\SigmaO$. 
After plugging (\ref{Eq:G-mat}) into (\ref{Eq:sigma-0}), 
one obtains the following integral equation:
\be
\SigmaO(P)=m^2-M_0^2+\delta m_{2,B}^2+
\frac{v^2}{2}\Lambda_2^{\vac}(0,P)
-i\frac{\lambda+\delta \lambda_0}{2} 
\int_K\Big(\Gv(K)\Big)^2\SigmaO(K)
+\frac{\lambda+\delta\lambda_0}{2}\int_K \Gr(K).
\label{Eq:Sigma0_def}
\ee
One can search for the solution of $\SigmaO$ in the following form: 
\be
\SigmaO(P)=\Gamma_m^{\vac}+\frac{v^2}{2}\Gamma_2^{\vac}(0,P)
+\frac12\Gamma^{\vac}\int_K \Gr(K).
\label{Eq:Sigma0}
\ee
Using (\ref{Eq:Sigma0}) in (\ref{Eq:Sigma0_def}) one finds that
$\Gamma_m^{\vac},$ $\Gamma_2^{\vac}(0,P),$ and $\Gamma^{\vac}$
fulfill the following Bethe-Salpeter-type equations:
\begin{subequations}
\bea
\Gamma_m^{\vac}&=&m^2-M_0^2+\delta m_{2,B}^2
-i\frac{\lambda+\delta \lambda_0}{2} \Gamma_m^{\vac}
\int_K \Big(\Gv(K)\Big)^2,
\label{Eq:BS-c}
\\
\Gamma_2^{\vac}(0,P)&=&\Lambda_2^{\vac}(0,P)
-i\frac{\lambda+\delta \lambda_0}{2}\int_K\Big(\Gv(K)\Big)^2
\Gamma_2^{\vac}(0,K), 
\label{Eq:BS-a}
\\
\Gamma^{\vac}&=&\lambda+\delta \lambda_0
-i\frac{\lambda+\delta \lambda_0}{2}\Gamma^{\vac}
\int_K \Big(\Gv(K)\Big)^2,
\label{Eq:BS-b}
\eea
\end{subequations}
in which all integrals are divergent. To render these divergent
equations finite, one has to impose some renormalization conditions
which will also determine the appropriate counterterms. Because of our
introduction of the $M_0$ scale and the splitting of the mass
counterterm into two pieces, we need one more renormalization condition
in addition to (\ref{Eq:ren_cond_sigma-vac}) to fix also 
$\delta m_{2,B}^2.$  By using (\ref{Eq:sigma-vac}) one sees that
$\displaystyle (m^2-M_0^2)\frac{\partial \Sigmav}{\partial M_0^2}$
satisfies \bea (m^2-M_0^2)\frac{\partial \Sigmav}{\partial M_0^2}
&=&m^2-M_0^2+(m^2-M_0^2)\frac{\partial (\delta m_{2,A}^2)}{\partial M_0^2}
-i\frac{\lambda+\delta \lambda_0}{2} 
(m^2-M_0^2) \frac{\partial \Sigmav}{\partial M_0^2}
\int_K \Big(\Gv(K)\Big)^2,
\eea
which is the same equation fulfilled by $\Gamma_m^{\vac}$, up to counterterms.
Since $\partial\Sigmav/\partial M_0^2=1,$
a simple renormalization condition on $\Gamma_m^{\vac}$ is
\be
\Gamma_m^{\vac}=m^2-M_0^2.
\label{Eq:gamma-m_cond}
\ee
This fixes the $\delta m_{2,B}^2$ counterterm:
\be
\delta m_{2,B}^2=(m^2-M_0^2)\frac{\partial (\delta m_{2,A}^2)}{\partial M_0^2}=
-\frac{\lambda+\delta\lambda_0}{2}(m^2-M_0^2)T_d^{(0)},
\label{Eq:delta-m22B}
\ee
where $T_d^{(0)}$ is defined in (\ref{Eq:Td2_Td0_S0}). 
From (\ref{Eq:delta-m22A}) and (\ref{Eq:delta-m22B}) 
one obtains the complete $\delta m_2^2$ counterterms, as 
the sum of $\delta m^2_{2,A}$ and  $\delta m^2_{2,B},$ with the 
expression given in (\ref{Eq:delta-m22}).
Before moving to the renormalization of (\ref{Eq:BS-a}) and
(\ref{Eq:BS-b}) we note that $\delta m_2^2$ can be equivalently fixed 
to the same value by keeping the condition (\ref{Eq:gamma-m_cond}) but  
replacing (\ref{Eq:ren_cond_sigma-vac}) with the condition
\be
\Sigma(P)\big|_{\textrm{overall}}=m^2,
\label{Eq:overall_cond1}
\ee 
where ``overall'' refers to the environment independent part of the
self-energy, which does not coincide with $\Sigma^{(v)}$ due to
additional terms coming from $\Sigma^{(0)},$ and contains the usual
overall divergence. One can identify this
part of the self-energy as the one which remains after $v$ and $\Gr$
are formally set to zero. Then, using the condition
(\ref{Eq:gamma-m_cond}) in the right-hand side of (\ref{Eq:BS-c}),
with the notation introduced in (\ref{Eq:Td2_Td0_S0}) one obtains in
(\ref{Eq:Sigma0}) 
$\SigmaO(P)\big|_{\textrm{vac}}=m^2-M_0^2+\delta m^2_{2,B}+ 
(\lambda+\delta\lambda_0)(m^2-M_0^2) T_d^{(0)}/2,$ 
so that upon adding it to $\Sigmav$ of (\ref{Eq:sigma-vac}), 
the condition (\ref{Eq:overall_cond1}) becomes
\be
m^2+\delta m_2^2+\frac{\lambda+\delta\lambda_0}{2} 
\left(T_d^{(2)}+(m^2-M_0^2) T_d^{(0)}\right)=m^2.
\ee

Next, we focus on the renormalization of $\Gamma_2^{\vac}$ and $\Gamma^{\vac}.$
As a renormalization condition we impose
\bea
\Gamma^{\vac}=\lambda,
\label{Eq:Gamma_cond}
\eea
which using (\ref{Eq:BS-b}) leads to
(\ref{Eq:delta-lambda-0}) for $\delta \lambda_0$. Then, using
(\ref{Eq:BS-a}) and (\ref{Eq:Lambda2_vac}), one sees that the
difference
\bea
\Gamma_2^{\vac}(0,P)-\Gamma_2^{\vac}(0,0)=-i\lambda^2\int_K 
\left[\Gv(K+P)\Gv(K)-\Big(\Gv(K)\Big)^2\right]
=\lambda^2 I_F^{\vac}(P)
\eea
is finite, where we defined $I_F^{\vac}(P)$ as the finite part of the
bubble integral in the zero-momentum subtraction scheme in which
$I_F^\vac(0)=0.$ This shows that one only needs to renormalize the
momentum independent $\Gamma_2^{\vac}(0,0)$ which is achieved by
imposing on it the renormalization condition
\be
\Gamma_2^{\vac}(0,0)=\lambda.
\label{Eq:Gamma2_cond}
\ee
Then, the explicitly finite $\Gamma_2^{\vac}(0,P)$ function is given by
\be
\Gamma_2^{\vac}(0,P)=\lambda+\lambda^2 I_F^{\vac}(P).
\label{Eq:fin-Gamma2}
\ee
Using (\ref{Eq:fin-Gamma2}) in (\ref{Eq:BS-a}) 
one obtains the equation determining $\delta \lambda_2$, that is 
(\ref{Eq:delta-lambda-2}).

In conclusion, the imposed renormalization conditions give explicitly
finite expressions for the functions $\Gamma_m^{\vac},$ $\Gamma_2^{\vac}(0,P),$
and $\Gamma^{\vac}$ appearing in (\ref{Eq:Sigma0}), and provide also 
the following finite equation for $\SigmaO(P)$: 
\be
\SigmaO(P)=m^2-M_0^2
+\frac{v^2}{2}\Big(\lambda+\lambda^2 I_F^{\vac}(P)\Big)
+\frac{\lambda}{2}\int_K \Gr(K).  
\label{Eq:Sigma0-final}
\ee

Two interesting remarks are in order at this point.
The first concerns the solution of (\ref{Eq:BS-a}) which, as one can check
iteratively, is easily expressed in terms of $\Gamma^{\vac}$ as
\be
\Gamma_2^{\vac}(0,P)=\Lambda_2^{(v)}(0,P)-\frac{i}{2}\Gamma^{\vac}
\int_K\Lambda_2^{(v)}(0,K)\Big(\Gv(K)\Big)^2.
\label{Eq:Gamma-tilde-2}
\ee
The second comment refers to obtaining a different expression for the 
$\delta\lambda_2$ counterterm. Using (\ref{Eq:Gamma_cond}), 
(\ref{Eq:Gamma2_cond}), and (\ref{Eq:fin-Gamma2}) in 
(\ref{Eq:Gamma-tilde-2}) one obtains
\be
0=\delta\lambda_2+\frac{1}{2}\lambda(3\lambda+\delta\lambda_2)T_d^{(0)}+
\frac{1}{2}\lambda^3\left((T_d^{(0)})^2+T_d^{(I)}\right).
\label{Eq:delta-lambda-2b}
\ee
Note that (\ref{Eq:delta-lambda-2b}) does not coincide with
(\ref{Eq:delta-lambda-2}), but the solutions for $\delta \lambda_2$
do. Indeed, using the result of (\ref{Eq:delta-lambda-0}) for $\delta
\lambda_0$ in (\ref{Eq:delta-lambda-2}), the same expression as that 
coming from (\ref{Eq:delta-lambda-2b}) is obtained. The Eq. 
(\ref{Eq:delta-lambda-2b}) served in \cite{patkos08} 
as a consistency check of the renormalization procedure.

\begin{figure}[t]
\begin{center}
\includegraphics[keepaspectratio,width=0.7\textwidth,angle=0]{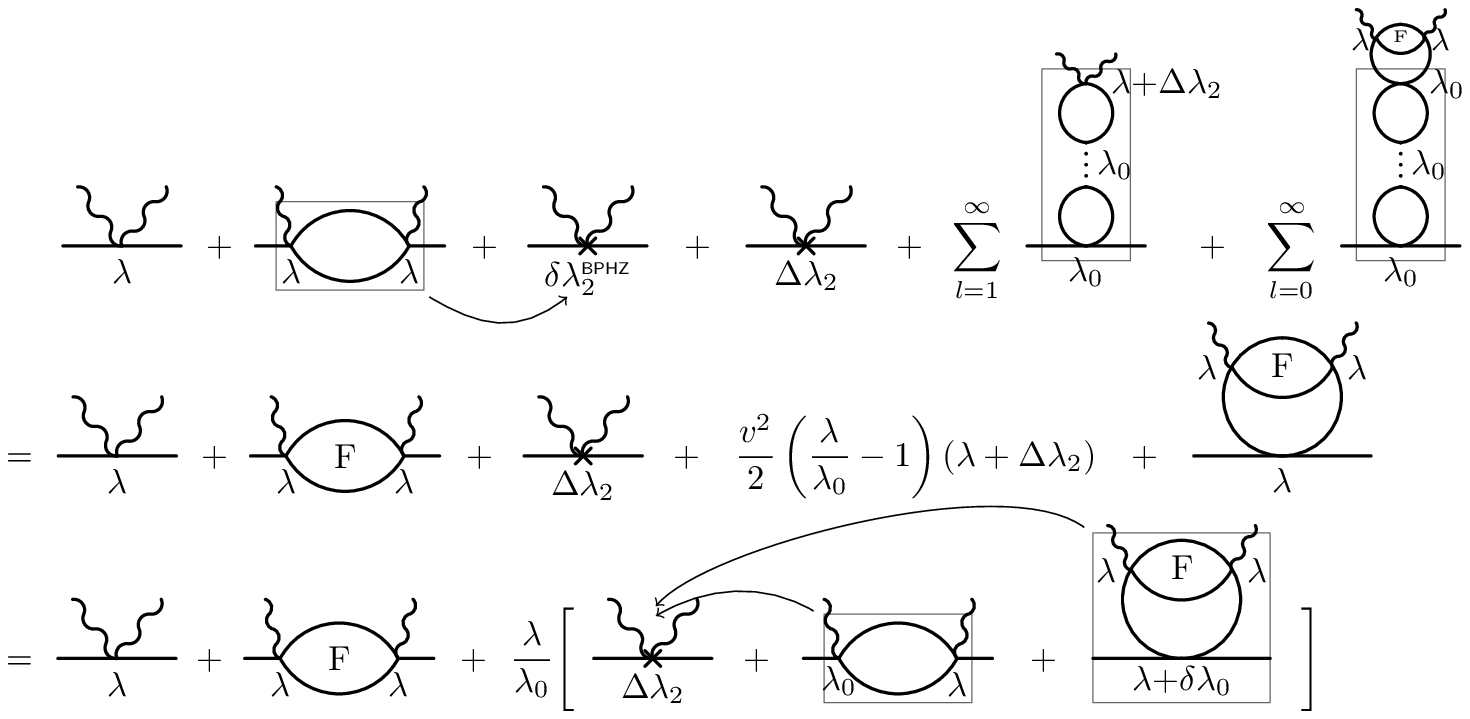}
\caption{Diagrammatic renormalization of the $v^2$-dependent part of
  the self-energy. See the text for details. The symmetry factors are
  not indicated and we used the shorthand
  $\lambda_0=\lambda+\delta\lambda_0.$ The boxes denote the divergence
  of the encircled part of the graph, the plain line denotes the
  propagator $\Gv(P)\equiv G_0(P),$ while ``F'' denotes the finite part
  of the bubble integral. The index $l$ in the sums goes over the
  possible number of bubbles (for the two classes of diagrams the
  lowest number of bubbles is zero or one, as indicated).}
\label{Fig:ren_with_cts}
\end{center}
\end{figure}

We close this subsection with a diagrammatic illustration of the
renormalization of the part proportional to $v^2$ in
Eq.~(\ref{Eq:sigma-0}) for $\Sigma^{(0)},$ see
Fig.~\ref{Fig:ren_with_cts}. This is interesting because it shows the
role the coupling counterterms play in the removal of both
subdivergences and overall divergences. The first line in the figure
shows the $v^2$-dependent diagrams generated upon iterating
Eq.~(\ref{Eq:sigma-0}) after the Bogoliubov-Parasiuk-Hepp-Zimmermann
(BPHZ) subtraction procedure was implemented
(see \cite{Berges:2005hc} for details.) The counterterm
$\delta\lambda_2$ displayed in Fig.~\ref{Fig:cts} is decomposed in the
sum of two terms $\delta\lambda_2^{\rm BPHZ}$ and $\Delta\lambda_2,$
and $\delta\lambda_2^{\rm BPHZ}$ is used to absorb the divergence of
the bubble integral present in $\Lambda_2^{(\rm v)}(0,P).$ The
remaining finite part of the bubble integral is denoted by 'F'.  In
order to obtain the second line of the figure we localized all the
subdivergences. Then, we used the renormalization condition
(\ref{Eq:Gamma_cond}). To obtain the third line of the figure, a
cancellation occuring between the third and fourth term of the second
line was exploited and the consequence of the renormalization
condition (\ref{Eq:Gamma_cond}), that is $\lambda-\lambda_0=(\lambda_0
\lambda/2) \int_K \Gv(K),$ where $\lambda_0=\lambda+\delta\lambda_0,$
was rewritten in a diagrammatic form.  The remaining overall
divergences of the diagrams in the third line are removed by
$\Delta\lambda_2.$ Since the divergent integrals associated with these
diagrams are $T_d^{(0)}$ and $T_d^{(I)},$ keeping in mind that one has
to add also $\delta\lambda_2^{\rm BPHZ},$ from the condition of
vanishing of the square bracket in the last line of
Fig.~\ref{Fig:ren_with_cts} one obtains the expression of
$\delta\lambda_2$ determined previously.

\subsection{Renormalization of the field equation in the broken symmetry phase}

We use in the field equation (\ref{Eq:EoS}) the splitting of the
propagator introduced in the previous subsection in (\ref{Eq:G_split})
and obtain
\bea
0&=&m^2+\delta m_0^2+\frac{\lambda+\delta\lambda_2}{2}\int_P \Gv(P)
-i\frac{\lambda^2}{6}\int_P\int_K \Gv(P+K) \Gv(P) \Gv(K)
\nonumber \\
&&+\frac{\lambda+\delta \lambda_4}{6}v^2
+\frac12\int_K\Lambda_2^{\vac}(0,K) \Gm(K)
-i\frac{\lambda^2}{6}\int_P\int_K\Big(3\Gv(P)+\Gm(P)\Big) 
\Gm(P+K)\Gm(K).
\eea
Using (\ref{Eq:G-mat}) and (\ref{Eq:Sigma0}) in the second line, we obtain
\bea
0&=&m^2+\delta m_0^2+\frac{\lambda+\delta \lambda_2}{2}\int_P \Gv(P)
-i\frac{\lambda^2}{6}\int_P \int_K \Gv(P) \Gv(P+K)\Gv(K)
\nonumber\\
&&-\frac{i}{2}\Gamma_m^{\vac}\int_K\Lambda_2^{\vac}(0,K) \Big(\Gv(K)\Big)^2
+\frac{v^2}{6}\Gamma_4^{\vac}
+\frac12 \int_K \Gamma_2^{\vac}(0,K) \Gr(K)
\nonumber\\&&
-i\frac{\lambda^2}{6}\int_P \int_K
\Big(3\Gv(P)+\Gm(P)\Big)\Gm(P+K)\Gm(K),
\label{Eq:EoS-2}
\eea
where we used the expression of $\Gamma_2^{\vac}$ given in 
(\ref{Eq:Gamma-tilde-2}) and defined
\be
\Gamma_4^{\vac}=\lambda+ \delta \lambda_4-\frac32 i\int_K
\Lambda_2(0,K)\Big(\Gv(K)\Big)^2 \Gamma_2^{\vac}(0,K). 
\label{Eq:Gamma-4}
\ee
The integral in the expression above is divergent, and one imposes the
following renormalization condition on $\Gamma_4^{\vac}:$ 
\bea
\Gamma_4^{\vac}=\lambda.
\eea
This leads to the equation for $\delta \lambda_4$ given in
(\ref{Eq:delta-lambda-4}). 

Since the last but one term of (\ref{Eq:EoS-2}) is already finite in
view of (\ref{Eq:Gamma2_cond}) and (\ref{Eq:fin-Gamma2}), and the last
term is a finite integral, we are left with the divergences of the
first line and the first term of the second line. In order to complete
the renormalization one only has to find the quantity on which the
renormalization condition determining $\delta m_0^2$ can be imposed.
It is straightforward to check that 
\be
\frac{\delta^2 V[v,G_v]}{\delta v \delta v}\bigg|_{v=0}=
m^2+\delta m_0^2+\frac{\lambda+\delta\lambda_2}{2}\int_P G_{v=0}(P)
-i\frac{\lambda^2}{6}\int_P\int_K G_{v=0}(P+K) G_{v=0}(P) G_{v=0}(K),
\ee 
where $G_v$ on the left-hand side is the solution of the stationarity
condition $\delta V[v,G_v]/\delta G_v=0$ and the subscript indicates
explicitly that it depends on the vacuum expectation value $v,$ 
so that the chain rule has to be applied when taking the derivatives
with respect to $v.$ With the splitting of the propagator $G_{v=0}$
into vacuum, matter, and ``regular'' parts, one can proceed
exactly as at the beginning of this subsection, only that one has to
omit everywhere the $v$-dependent terms. Then, one has
\bea
\frac{\delta^2 V[v,G_v]}{\delta v \delta v}\bigg|_{v=0}
&=&m^2+\delta m_0^2+\frac{\lambda+\delta \lambda_2}{2}\int_P \Gv(P)
-i\frac{\lambda^2}{6}\int_P \int_K \Gv(P) \Gv(P+K)\Gv(K)
\nonumber\\
&&-\frac{i}{2}\Gamma_m^{\vac}\int_K\Lambda_2^{\vac}(0,K) \Big(\Gv(K)\Big)^2
+\frac12 \int_K \Gamma_2^{\vac}(0,K) \Gr_{v=0}(K)
\nonumber\\
&&-i\frac{\lambda^2}{6}\int_P \int_K
\Big(3\Gv_{v=0}(P)+\Gm_{v=0}(P)\Big)\Gm_{v=0}(P+K)\Gm_{v=0}(K).
\label{Eq:magyarazat}
\eea 
The last term above is finite and since $\Gamma_2^{\vac}(0,K)$ was
already made finite by imposing the condition (\ref{Eq:Gamma2_cond}),
the last but one term above is also finite.  Therefore, one could in
principle formulate our last renormalization condition as
$\frac{\delta^2 V[v,G_v]}{\delta v \delta v} \big|_\textrm{v=0}=m^2,$
but then $\delta m_0^2$ determined through this condition would differ
by a finite term from $\delta m_0^2$ given in (\ref{Eq:delta-m02}).
In order to obtain the desired expression for $\delta m_0^2$ one
choose the following condition
\be
\frac{\delta^2 V[v,G_v]}{\delta v \delta v} \bigg|_\textrm{v=0, overall}
=m^2,
\label{Eq:overall_cond2}
\ee 
where the concrete expression on the left-hand side is given by the
right-hand side of (\ref{Eq:magyarazat}) without the last two terms,
and the overall notation indicates that the divergence of
this term is the usual overall divergence, that is the divergence
which remains after all subdivergences are removed.
\footnote{The quantity on the left-hand
side of (\ref{Eq:overall_cond2}) is obtained formally by setting 
$v$ and $\Gr$ to zero after (\ref{Eq:G_split}), (\ref{Eq:G-mat}),
and (\ref{Eq:Sigma0_def}), (\ref{Eq:Sigma0}) 
are used in the explicit expression of 
$\frac{\delta^2 V[v,G_v]}{\delta v \delta v}.$ }
Then, it is easy to check that (\ref{Eq:overall_cond2}) gives
indeed for $\delta m_0^2$ the expression in (\ref{Eq:delta-m02}).  

The two renormalization conditions (\ref{Eq:Gamma-4}) and 
(\ref{Eq:delta-m02})
together with the expression of $\Gamma_2^{\vac}(0,K)$ given in
(\ref{Eq:fin-Gamma2}) leads to the explicitly finite version of the
field equation 
\bea 0&=&m^2+\frac{\lambda}{6}v^2+\frac{\lambda}{2}
\int_P \Big(1+\lambda I_F^{\vac}(P)\Big)\Gr(P)
\nonumber\\ &&-i\frac{\lambda^2}{2}\int_P\int_K \Gv(P+K)\Gm(P)\Gm(K)
-i\frac{\lambda^2}{6}\int_P \int_K \Gm(P+K)\Gm(P)\Gm(K).
\label{Eq:EoS-fin}
\eea

The renormalization conditions imposed above give exactly the same
expressions for these counterterms introduced in the functional
(\ref{Eq:l2-2PI-Vct}) as those appearing in \cite{patkos08} (see also
[\ref{Eq:all_counterterms})]. This demonstrates the equivalence
between the two renormalization procedures. We note that even though
we formulated our set of renormalization conditions on the quantities
$\Gamma^{\vac}, \Gamma_2^{\vac}(0,P),$ $\Gamma_4^{\vac}$,
corresponding to the three four-point functions $\bar V,$
$V(0,P)=\delta^2 \Sigma(P)/\delta v^2,$ $\hat V=\delta^4
V[v,G_v]/\delta v^4,$ introduced in \cite{Berges:2005hc}, and also
using the curvature $\delta^2 V[v,G_v]/\delta v\delta v$, they do not
look as natural as those imposed in
\cite{Berges:2005hc,arrizabalaga06} at $T=T^*$ and in the symmetric
phase of the model ($v=0$).  In our case $\Gv$ is not the full
propagator at $v=0$ and in order to obtain a given expression of the
counterterms which contains $\Gv$ in place of the full propagator we
need a specific way to formulate the renormalization condition in
(\ref{Eq:overall_cond1}) and (\ref{Eq:overall_cond2}).  One may even
say that some of our conditions used for renormalization are not
genuine renormalization conditions because they are not formulated on
quantities accessible from the effective potential through functional
derivatives and resemble more the way divergences are removed in a
minimal subtraction scheme. As we have already formulated, we imposed
these conditions in order to facilitate the comparison, to be done in
the next section, between the solution of the finite equations and
those using counterterms. On the other hand, as discussed in
Ref.~\cite{Reinosa:2011ut}, finding a set of natural renormalization
condition, prescribing e.g. the mass defined from the
  variational propagator and the curvature at the minimum of the
  effective potential\footnote{This two quantities do not necessarily
    coincide in a given truncation of the 2PI effective potential.}
together with the value of the independent four-point functions is
problematic when renormalizing the model at $T=0,$ that is in the
broken symmetry phase, even at the Hartree level approximation of the
2PI effective potential.

\section{Numerical implementation and results \label{sec:implement}}
Before discussing the algorithms and presenting the results we make
some general statements about the numerical method used. 

Since we solve iteratively the set of coupled integral equations
consisting of the self-consistent propagator equation and the field
equation, one has to store and upgrade in each iteration step the
value of some functions on a grid and to approximate using
interpolation the value of these functions at the points required by
the integration routine.  For this purpose we use an equidistant grid
for the modulus of the four-momentum, the one-dimensional Akima spline
interpolation method and the numerical integration routines of the GNU
Scientific Library \cite{gsl}. When we solve the set of finite
equations, one stores the regular and the matter part of the
propagator [see (\ref{Eq:G-mat})], while when solving the set of
equations with the counterterms it is the full self-energy which is
stored on the grid [see (\ref{Eq:sigma_decomp} and
(\ref{Eq:sigmas})]. In the first case the integrals are convergent in
the UV, but nevertheless, for the functions which are stored on the
grid we have to take into account the fact that there is a maximal
value of the modulus of the four-momentum. The convolution of these
functions should be calculated accordingly, using $\theta$ functions
to restrict the momenta of the propagators, see (\ref{Eq:iden1}). For
functions of the momentum with a known analytical expression there is
no need to store them on the grid, and in consequence a simplification
is encountered in the convolution involving such functions, see
e.g. (\ref{Eq:I_mv}). In contradistinction to this, when counterterms
are used, all momentum-dependent functions are cut at the maximal
value of the modulus of the four-momenta, that is at the physical
cutoff $\Lambda,$ irrespective of the fact that they are stored or not
stored on the grid, in the latter case their expression is known
analytically. We mention here that cutting the sum of momenta in the
case of the bubble integral and as a consequence in all the double
integrals encountered seems to be the cleanest way to
proceed. \footnote{See \cite{urko11} for a discussion on the way such
  a regularization can be implemented at the level of the
  path-integral.} In fact, in the case of solving the set of finite
equations we could afford to simplify the calculation, by cutting only
the loop momenta in the integrals involving the propagator
$\Gv(P)\equiv G_0(P)$ which is not stored on the grid, because the
integrals are all convergent in the UV, and in consequence the
corrections are suppressed. We will see indeed that the solution of
the finite equations nicely agrees with that obtained by solving the
equations with counterterms. Moreover, cutting only the loop-momenta,
when solving the set of equation containing the counterterms, produces
some small differences compared to the case when the sum of momenta is
also cut, at least up to the largest cutoff we investigated, and in
consequence is not a viable method for obtaining accurate numerical
results.

\subsection{Algorithm for solving the finite propagator and field equations
\label{ss:Euclidean-algorithm}}

We will solve in Euclidean space the explicitly finite field and
propagator equations obtained in the previous section. 
A Wick rotation is performed in every integral appearing in
our equations: for example with the continuation 
$k_0\to i k^4$ the tadpole integral reads 
$\int_K G(k_0,\k)\to i\int_{k_E} G(i k^4,\k)=\int_{k_E} \Delta(k_E)$
where the Euclidean four-momenta is $k_E=(\k,k^4)$, such that 
$k_E^2=\k^2 + k_4^2$ and the Euclidean propagator is defined as
$\Delta(k_E)=1/(k_E^2+\Sigma(k_E)).$

Using the expansion (\ref{Eq:prop}) of the propagator around the
vacuum propagator $\Dv(k_E)=1/(k_E^2+M_0^2)$ and the 
definition (\ref{Eq:G-mat}), the matter and regular parts of the 
Euclidean propagator reads
\bea
\Dm(k_E)&=&\frac{1}{k_E^2+M_0^2+\SigmaO(k_E)
+\Sigmar(k_E)}-\frac{1}{k_E^2+M_0^2},
\label{Eq:eucl-prop-m}
\\
\Dr(k_E)&=&\Dm(k_E)+\frac{\SigmaO(k_E)}{(k_E^2+M_0^2)^2}.
\label{Eq:eucl-prop-r}
\eea
The two pieces of the self-energy which appear above are given by 
\bea
\SigmaO(k_E)&=&m^2-M_0^2+\lambda\Big(1+\lambda I_F^{\vac}(k_E)\Big)\frac{v^2}{2}
+\frac{\lambda}{2}\int_{k_E}\Dr(k_E), 
\label{Eq:sigma0-eucl}
\\
\Sigmar(k_E)&=&
\lambda^2 v^2 I^\textrm{(mv)}(k_E)+\frac{\lambda^2v^2}{2} I^\textrm{(mm)}(k_E),
\label{Eq:sigmar-eucl}
\eea
where the ``matter-vacuum'' and ``matter-matter'' bubble integrals are
defined as
\begin{subequations}
\bea
I^\textrm{(mv)}(k_E)&=&-\int_{p_E}\Dm(p_E)\Dv(p_E+k_E),\\
I^\textrm{(mm)}(k_E)&=&-\int_{p_E}\Dm(p_E)\Dm(p_E+k_E).
\eea
\label{Eq:Imv_Imm}
\end{subequations}
The finite part of the vacuum bubble integral, $I_F^{\vac}$, appearing
in (\ref{Eq:sigma0-eucl}), is defined in (\ref{Eq:fin-bubble}).
After analytical continuation to Euclidean space it has the following
expression (recall that $G_0\equiv \Gv$):
\be
I_F^{\vac}(k_E)=\frac{1}{16\pi^2}\Bigg[\sqrt{1+\frac{4M_0^2}{k_E^2}}
\ln\frac{\sqrt{1+\frac{4M_0^2}{k_E^2}}+1}
{\sqrt{1+\frac{4M_0^2}{k_E^2}}-1}-2\Bigg].
\label{Eq:IF_eucl}
\ee
Note that a self-consistent equation for the regular part $\Dr$ can be
obtained in principle if one substitutes the expression of $\Dm$ from
(\ref{Eq:eucl-prop-m}) into (\ref{Eq:eucl-prop-r}) because $\SigmaO$
contains the integral of $\Dr$ and $\Dm$ appearing in the integrals of
$\Sigmar$ can be expressed in terms of $\Dr$ and $\SigmaO.$
Nevertheless, we store on the grid two quantities, $\Dm$ and $\Dr,$
which have to be solved simultaneously with the field equation
(\ref{Eq:EoS-fin}), which in Euclidean space reads
\be
0=m^2+\frac{\lambda}{6}v^2+\frac{\lambda}{2}\int_{k_E} 
\Big(1+\lambda I_F^{\vac}(k_E)\Big)\Dr(k_E)+
\frac{\lambda^2}{6}\big(3 S^\textrm{(mmv)} + S^\textrm{(mmm)}\big),\\
\label{Eq:EoS-fin-eucl}
\ee
where the corresponding pieces of the setting-sun integral are defined
as
\begin{subequations}
\bea
S^\textrm{(mmv)}&=&-\int_{p_E}\int_{k_E}\Dm(p_E)
\Dm(k_E) \Dv(p_E+k_E), \\
S^\textrm{(mmm)}&=&-\int_{p_E}\int_{k_E} 
\Dm(p_E)\Dm(k_E) \Dm(p_E+k_E).
\eea
\end{subequations}
Our iterative algorithm for solving the coupled set of equations for
$\Dr,$ $\Dm,$ and $v$ goes as follows. One starts at {\it zeroth}
order by neglecting all the integrals in the self-energy and the field
equation, including the finite part of the bubble $I_F^\vac.$ By doing
so, from (\ref{Eq:EoS-fin-eucl}), (\ref{Eq:eucl-prop-m}), and
(\ref{Eq:eucl-prop-r}) one obtains:
\[
v_0=\sqrt{-\frac{6m^2}{\lambda}},\ \ \quad 
\Dm_0(k_E)=-\frac{m^2+\frac{\lambda}{2}v^2-M_0^2}{(k_E^2+M_0^2)(k_E^2+m^2+\frac{\lambda}{2}v^2)},\ \ \quad
\Dr_0(k_E)=\frac{\left(m^2+\frac{\lambda}{2}v^2-M_0^2\right)^2}{(k_E^2+M_0^2)^2(k_E^2+m^2+\frac{\lambda}{2}v^2)}.
\]
Then, starting from the {\it first} order, the iteration is done in two
steps.  At a generic order $n\ge 1$ one first upgrades on the grid
$\Dm_n$ and $\Dr_n$ using (\ref{Eq:eucl-prop-m}) and
(\ref{Eq:eucl-prop-r}), respectively, where the self-energy, which
includes now $I_F^\vac,$ is calculated with the lower order
quantities: $v_{n-1},$ $\Dm_{n-1},$ $\Dr_{n-1}.$ As a second step of
the n{\it th} iteration, one upgrades the vacuum expectation value by
calculating $v_{n}$ from (\ref{Eq:EoS-fin-eucl}) with the integrals
done with the already upgraded $\Dm_n$ and $\Dr_n.$ This procedure is
repeated until the iteration converges. The angular integration in the
integrals appearing in (\ref{Eq:sigmar-eucl}) and
(\ref{Eq:EoS-fin-eucl}) are performed in
Appendix~\ref{app:quantum-corr} as outlined at the beginning of this
section.

\subsection{Algorithm for solving the propagator and field equations using counterterms}

If one decides to solve the propagator and field equations using
counterterms, whose role in this equations is to cancel the divergent
part of the integrals, then one needs a method to determine them. Such
a method was developed in \cite{patkos08}, and we review it here
because it will be slightly modified for numerical reasons.  In terms
of the remaining finite part of the tadpole $T_{\rm F}[G]$ and bubble
integral $I_{\rm F}(P,G)$ the propagator equation (\ref{Eq:full_prop})
reads
\be
i G^{-1}(P)=P^2-M^2-\frac{1}{2}\lambda^2 v^2 I_{\rm F}(P,G), \qquad
M^2=m^2+\frac{\lambda}{2}v^2+\frac{\lambda}{2}T_{\rm F}[G].
\ee
Using the expansion of $G$ around $G_0$
\be
G(P)=G_0(P)-i G_0^2(P)\left(
M^2-M_0^2+\frac{1}{2}\lambda^2 v^2 I_{\rm F}(P,G)\right)+\dots \,,
\ee
the tadpole, bubble, and setting-sun integrals can be explicitly
decomposed into divergent and finite parts (see Ref.~\cite{patkos08} 
for details):
\begin{subequations}
\bea
I(P,G)&=&T_d^{(0)}+I_{\rm F}(P,G),
\label{Eq:I_decomp}
\\
T[G]&=&T_d^{(2)}+(M^2-M_0^2) T_d^{(0)}+\frac{1}{2}\lambda^2 v^2 T_d^{(I)}
+T_{\rm F}[G],
\label{Eq:T_decomp}
\\
S(0,G)&=&S_0(0)+3 T_d^{(0)} T_{\rm F}[G]+3 (M^2 - M_0^2)\left(
\big(T_d^{(0)}\big)^2+T_d^{(I)}\right)
+\frac{3}{2}\lambda^2 v^2 \left(T_d^{(I)} T_d^{(0)}+T_d^{(I,2)}\right)
+S_{\rm F}(0,G),
\label{Eq:SS_decomp}
\eea
\label{Eq:TBSS_decomp}
\end{subequations}
\vspace{-2mm}   

\noindent
where the divergent integrals
$T_d^{(2)},T_d^{(0)},T_d^{(I)},T_d^{(I,2)}$ and $S_0(0)$ are defined
and calculated in Appendix~\ref{app:cutoff}.  Plugging the decomposed
expressions of the integrals (\ref{Eq:TBSS_decomp}) into the
propagator and field equations and subtracting from them the
corresponding explicitly finite equations, written in terms of the
renormalized parameters $m^2,\lambda$ and the finite part of the
integrals, one obtains relations between the counterterms and the
divergent integrals. Requiring that the coefficient of the $v^2,$
$v^0,$ and $T_F[G]$ vanish independently, one obtains the counterterms
determined in \cite{patkos08}. These were also obtained from
appropriate renormalization conditions in
Sec.~\ref{sec:finite_equation} and are summarized in
Appendix~\ref{app:cutoff}.
\footnote{Actually, applying the outlined procedure to the field
  equation one obtains (\ref{Eq:delta-lambda-2b}) from the
  cancellation of the coefficient of $v^2,$ but as stated below
  (\ref{Eq:delta-lambda-2b}), its solution for $\delta\lambda_2$
  coincides with the one obtained from the propagator equation, as it
  should.}

In Euclidean space these counterterms are functions of the four-dimensional
rotational invariant cutoff $\Lambda$ appearing through the integrals
explicitly calculated in (\ref{Eq:div_ints_euclid}).  They can be
used to solve iteratively the set of Euclidean equations
\begin{subequations}
\bea
\label{eq:Sigma}
\Sigma(p)&=&m^2+\delta m_2^2+\frac12(\lambda+\delta \lambda_2)v^2
+\frac12(\lambda+\delta \lambda_0)T(\Sigma)+
\frac12 \lambda^2 v^2 I(p,\Sigma), \\
\label{eq:EoS}
0&=&m^2+\delta m_0^2+\frac16(\lambda+\delta \lambda_4)v^2+
\frac12(\lambda+\delta \lambda_2)T(\Sigma)+\frac{\lambda^2}{6}S(0,\Sigma).
\eea
\end{subequations}
However, since the counterterms determined from
(\ref{Eq:all_counterterms}) are designed to cancel in these equations
the divergences of integrals produced by the full (i.e. iteratively
converged) propagator, they cancel the complete cutoff dependence of
$T[\Sigma],$ $I(P,\Sigma)$ and $S(0,\Sigma)$ only when the solution of
this set of equations has converged. This is not a problem in itself,
since, as it will be demonstrated, the iterative procedure converges,
but the convergence of the solution of the set of equations is rather
slow. It is possible to improve the iterative procedure if, instead of
using the counterterms given in (\ref{Eq:all_counterterms}), we
rederive them at each order of the iteration using the procedure
outlined above.  These counterterms, which come from the requirement
to have finite equations at each order of the iteration, will evolve
during the process of iteration toward the value of the counterterms
obtained from (\ref{Eq:all_counterterms}) which will be reached when
the iteration converged.  Since, as we will see, the new counterterms
are now environment dependent, that is depend on $v$ and
$T_F[\Sigma]$, this convergence can only happen if $v$ and $\Sigma$
also converge to their appropriate values, and has to be
checked numerically.

In what follows we present the improved iterative procedure which uses
evolving counterterms. At each order the iteration starts by the
upgrade of the self-energy, followed by the upgrade of the vacuum 
expectation value determined from the field equation.

At {\it zeroth} order of the iteration the field and the self-energy
are the tree-level ones:
\bea
v_0^2=-\frac{6m^2}{\lambda},\qquad  
\Sigma_0(p)=m^2+\frac{\lambda}{2}v_0^2.
\eea
At this order, since there are no quantum fluctuations at all, there
are no divergences to cancel and in consequence all the counterterms 
are zero, that is
$\delta m_{0,0}^2=\delta m_{2,0}^2=\delta\lambda_{0,0}=\delta\lambda_{2,0}=
\delta\lambda_{4,0}=0,$
where the counterterms also carry the index of the iteration number
because all of them will change during the iteration process.

The general formulas for the n{\it th} order counterterms to be given
below encompass all orders with the following prescription: at
$n=0$ one starts from the tree-level values, and formally
$T(\Sigma_n)=I(\Sigma_n)=0$ for $n\in\{-1,-2\}$ and $v_{-1}=v_0.$ Suppose that
the (n-1){\it th} order of the iteration is done, i.e. $v_{n-1}^2$ and
$\Sigma_{n-1}(p)$ together with the corresponding counterterms are
already known, then at n{\it th} order we start by upgrading the
self-energy using [c.f. (\ref{eq:Sigma})]:
\bea
\label{Eq:Sigma-itern}
\Sigma_n(p)=m^2+\delta m_{2,n}^2+\frac{\lambda}{2}v_{n-1}^2+\frac{\delta \lambda_{2,n}}{2}v_{n-2}^2+\frac{\lambda+\delta \lambda_{0,n}}{2}T(\Sigma_{n-1})+\frac{\lambda^2}{2}v_{n-1}^2 I(p,\Sigma_{n-1}).
\eea
The finite part of the tadpole and bubble are defined through
\bea
T(\Sigma_{n-1})&=&T_d^{(2)}+\Big(m^2+\frac{\lambda}{2}v_{n-2}^2+\frac{\lambda}{2}T_{F}(\Sigma_{n-2})-M_0^2\Big)T_d^{(0)}+\frac{\lambda^2 v_{n-2}^2}{2}T_d^{(I)}+T_F(\Sigma_{n-1}), 
\label{eq:tad_n-1}
\\
I(p,\Sigma_{n-1})&=&T_d^{(0)}+I_F(p,\Sigma_{n-1}).
\eea
The renormalization procedure requires the independent vanishing of the 
overall divergence, field-dependent subdivergence and $T_F$-dependent 
subdivergence:
\begin{subequations}
\label{Eq:cts-nb}
\bea
0&=&\delta m_{2,n}^2+\frac{\lambda+\delta\lambda_{0,n}}{2}\Big(T_d^{(2)}+(m^2-M_0^2)T_d^{(0)}\Big),\\
0&=&\delta \lambda_{2,n}
+\frac{\lambda}{2}T_d^{(0)}(\lambda+\delta \lambda_{0,n})
+\frac{\lambda^2}{2} T_d^{(I)} (\lambda+\delta \lambda_{0,n}) (1-\delta_{n,1})
+\lambda^2 T_d^{(0)}\frac{v_{n-1}^2}{v_{n-2}^2},\\
0&=&(\lambda+\delta \lambda_{0,n})\frac{\lambda}{2}T_d^{(0)}T_{F}(\Sigma_{n-2})
+\delta \lambda_{0,n}T_F(\Sigma_{n-1}).
\eea
\end{subequations}

The counterterms $\delta m_{2,n}^2$, $\delta \lambda_{2,n}$ and
$\delta \lambda_{0,n}$ are obtained from (\ref{Eq:cts-nb}), so that
$\Sigma_n(p)$ is finite and can be calculated. Note that for $n=1$ the
second term in the equation for $\delta\lambda_{2,n}$ vanishes as
it should because in this case there is no divergence due to the
bubble, that is there is no $T_d^{(I)}$ in $T(\Sigma_0)$, whose
expression is
\be
T(\Sigma_0)=T_d^{(2)}+\Big(m^2+\frac{\lambda}{2}v_0^2-M_0^2\Big)T_d^{(0)}+T_F(\Sigma_0).
\ee
Note also that for $n=1$ one has $\delta\lambda_{0,1}=0$, since 
$T_F(\Sigma_{-1})=0$ by convention.

We continue with the determination of $v_n^2$ from the field equation,
which at n{\it th} order of the iteration reads
\bea
\label{Eq:EoS-iter-n}
0&=&m^2+\delta m_{0,n}^2+\frac{\lambda}{6}v_n^2+\frac{\delta \lambda_{4,n}}{6}v_{n-1}^2+\frac{\lambda+\delta \tilde\lambda_{2,n}}{6}T(\Sigma_{n})+\frac{\lambda^2}{6}S(0,\Sigma_{n}),
\eea
where $T(\Sigma_n)$ is given by (\ref{eq:tad_n-1}) with $n-1\to n$ and
the setting sun is
\bea
S(0,\Sigma_n)&=&S_0(0)+3 T_d^{(0)} T_{\rm F}[\Sigma_n]+3 \Big(m^2 
+\frac{\lambda}{2}v_{n-1}^2+\frac{\lambda}{2}T_{F}(\Sigma_{n-1})-M_0^2\Big)
\left(\big(T_d^{(0)}\big)^2+T_d^{(I)}\right)\nonumber\\
&&+\frac{3}{2}\lambda^2 v^2_{n-1} \left(T_d^{(I)} T_d^{(0)}+T_d^{(I,2)}\right)
+S_{\rm F}(0,\Sigma_n).
\eea
Applying the same requirements as for the self-energy, one obtains:
\begin{subequations}
\label{Eq:ct-iter-n}
\bea
0&=&\delta m_{0,n}^2+\frac{\lambda+\delta \tilde\lambda_{2,n}}{2}\Big(T_d^{(2)}+(m^2-M_0^2)T_d^{(0)}\Big)+\frac{\lambda^2}{2}(m^2-M_0^2)\Big((T_d^{(0)})^2+T_d^{(I)}\Big)+\frac{\lambda^2}{6}S_{0}(0),\quad \\
0&=&\delta \lambda_{4,n}+\frac{3\lambda}{2}(\lambda+\delta \tilde\lambda_{2,n}+\lambda^2 T_d^{(0)})\Big(T_d^{(0)}+\lambda T_d^{(I)}\Big)+\frac{3\lambda^3}{2}\Big(T_d^{(I)}+\lambda T_d^{(I,2)}\Big), \\
0&=&\delta \tilde\lambda_{2,n}+\lambda^2 T_d^{(0)}+\frac{T_{F}(\Sigma_{n-1})}{T_F(\Sigma_{n})}\Big[\frac{\lambda}{2}(\lambda+\delta \tilde\lambda_{2,n})T_d^{(0)}+\frac{\lambda^3}{2}\Big((T_d^{(0)})^2+T_d^{(I)}\Big)\Big].
\eea
\end{subequations}
Note that a new counterterm $\delta \tilde{\lambda}_2$ has been
introduced instead of $\delta \lambda_2$, since in the $n$th iteration
$\delta \lambda_{2,n}$ does not eliminate properly the appropriate
subdivergences of the field equation.  Extracting $\delta m_{0,n}^2$,
$\delta \lambda_{4,n}$ and $\delta\tilde \lambda_{2,n}$, from
(\ref{Eq:ct-iter-n}) one can obtain $v_n^2$ from
(\ref{Eq:EoS-iter-n}). The counterterm $\delta\tilde \lambda_{2,n}$
will be equal to $\delta\lambda_{2,n}$ determined from the propagator
equation only when the iteration converged, so that
$T_F(\Sigma_{n-1})/T_F(\Sigma_{n})\to 1.$ In this limit and when
$v_n/v_{n-1}\to 1$ the counterterms will converge to the values
obtained from (\ref{Eq:all_counterterms}). This happens only
asymptotically and in practice the iteration stops when the relative
change in the values of the field and the self-energy are smaller than
some given value dictated by our requirement of accuracy. The same
iterative procedure is used when the counterterms are not evolved
during the iteration, but used as determined from
(\ref{Eq:all_counterterms}): one first upgrades the self-energy and
then the field equation and checks for their convergence.

\subsection{Numerical results}

\begin{figure}[!t]
\begin{center}
\raisebox{0.35cm}{
\includegraphics[keepaspectratio,width=0.465\textwidth,angle=0]{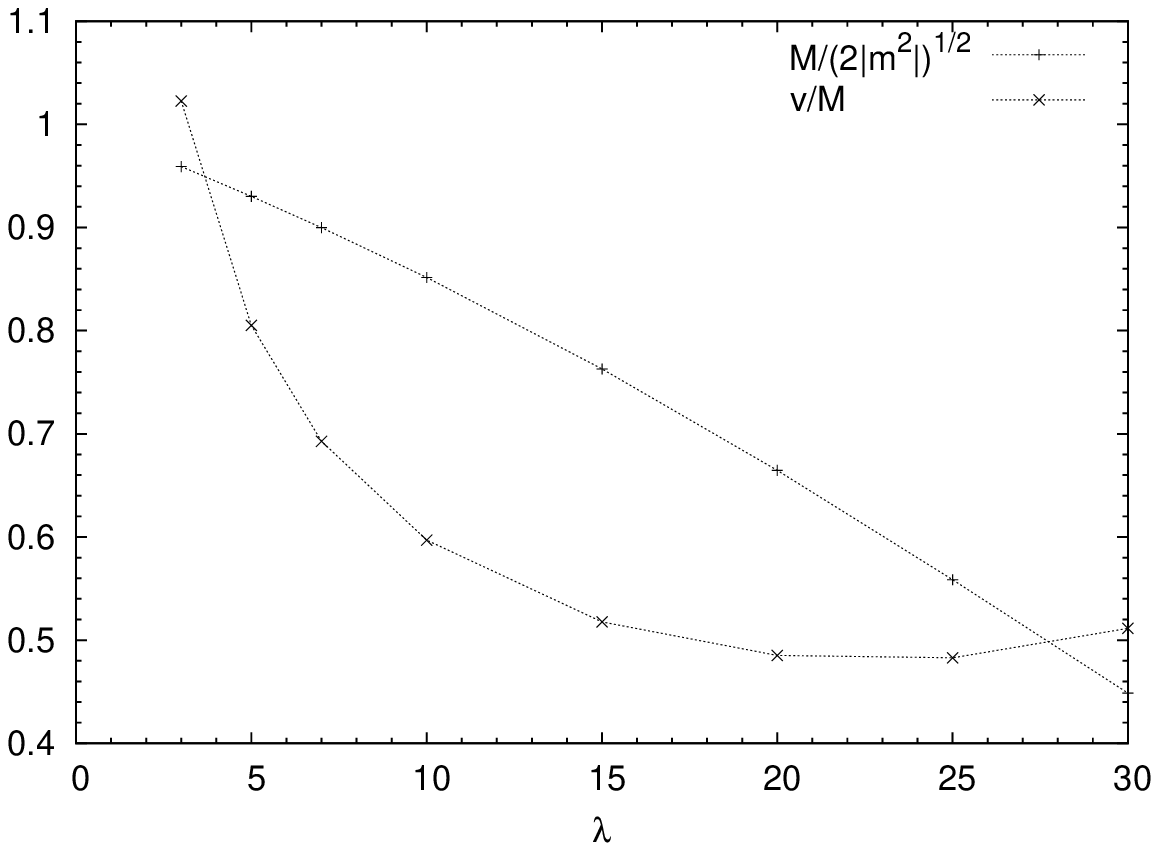}}
\includegraphics[keepaspectratio,width=0.495\textwidth,angle=0]{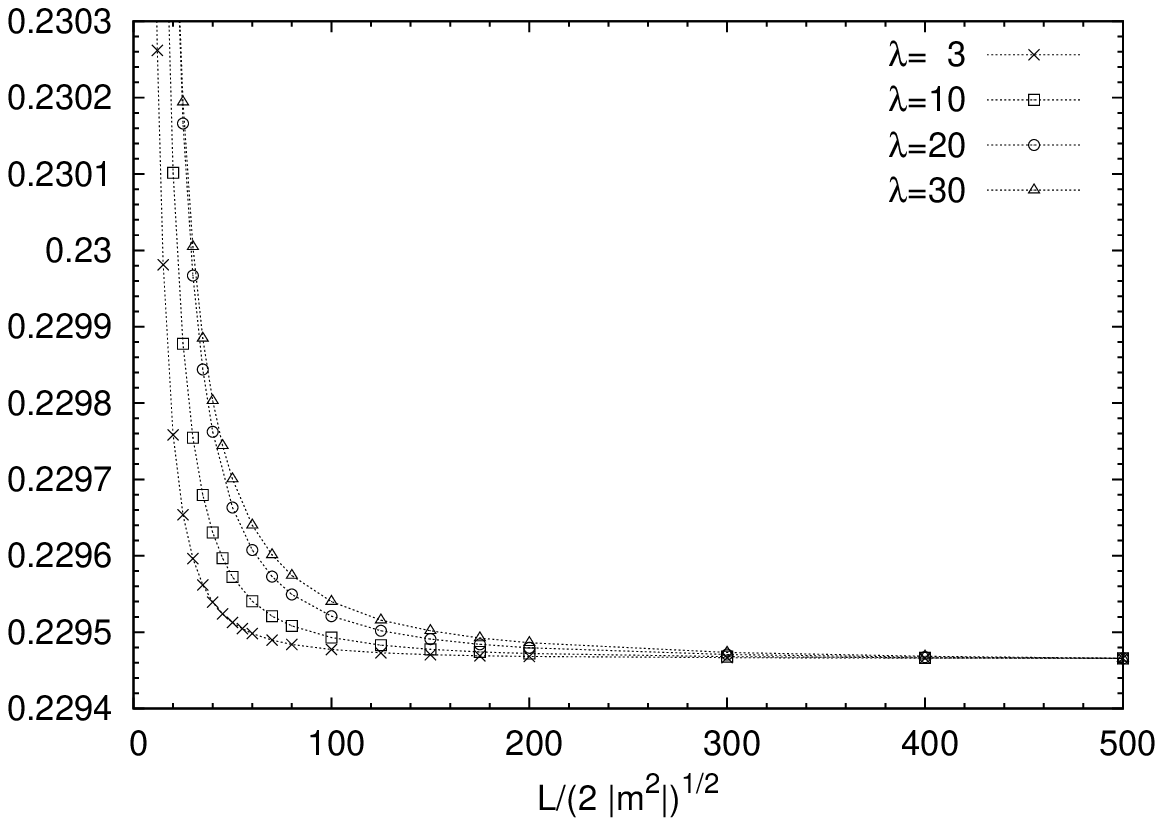}
\caption{The solution of the explicitly finite propagator and field
  equations.  Left panel: $v/M$ and $M=\Sigma^{\frac{1}{2}}(p=0)$ as
  function of the coupling obtained for $L/\sqrt{2|m^2|}=500$.  Right
  panel: The converged value of $v$ as function of the maximal value
  of the modulus of the momentum stored on the grid for different
  values of the coupling constant $\lambda$.  An appropriate constant
  was subtracted from the values of $v$ obtained for $\lambda=3,$
  $10,$ and $20$ such that the resulting values coincide at
  $L/\sqrt{2|m^2|}=500$ with the value of $v$ obtained for
  $\lambda=30.$
\label{fig:L_dep}}
\end{center}
\end{figure}

In what follows we present the results on the numerical solution of
the propagator and field equations in Euclidean space. When solving
the explicitly finite set of equations the maximal value of the
modulus of the four-momentum stored on the grid is
$L/\sqrt{2|m^2|}=500,$ while in the case of solving equations with
counterterms the highest value of the modulus stored, that is the
four-dimensional physical cutoff, is $\Lambda/\sqrt{2|m^2|}=200.$
The value of the mass scale is chosen to be $M_0/\sqrt{2|m^2|}=2.1$
and the value of the renormalized mass parameter squared is
$m^2=-0.5,$ in some arbitrary physical units. Actually, to have an
idea on the physical scale, one can require to have
$v/\Sigma^\frac{1}{2}(p=0)\simeq 0.66,$ which is
the value of the ratio $f_\pi/m_\pi$ of the pion decay constant to 
the pion mass for $f_\pi=93$~MeV and $m_\pi=140$~MeV, and
$\Sigma^\frac{1}{2}(p=0)=m_\pi.$ Then, from the left panel of
Fig.~\ref{fig:L_dep} one sees that the first condition can be meet by
choosing $\lambda\simeq 8,$ for which value the second condition means
$\sqrt{2|m^2|}\simeq 158$~MeV. In units for which $m^2=-0.5$ the step
size on the grid in general was $0.01.$ We have checked that by
halving the step size the change in the results is comparable with or
smaller than the precision required by our convergence criterion. The
iteration was stopped when the change in the vacuum expectation value
was smaller than $10^{-6}.$ The precision of the numerical integration
routines used was higher than that, since we required relative error
bounds between $10^{-6}$ and $10^{-8}$.

The solution of the explicitly finite equations can be seen in
Fig.~\ref{fig:L_dep}. In the right panel one can see the converged
solution of the field equation obtained for various values of the
coupling as function of $L,$ the maximal value of the modulus stored
on the grid.  As explained in the figure caption, $v$ was shifted
appropriately in order to make all four curves meet at
$L/\sqrt{2|m^2|}=500.$ In all cases a plateau can be observed for
increasing values of $L.$ However, with increasing value of the
coupling $\lambda$ the plateau starts at a higher $L.$ With the same
convergence criterion the number of iterations increases with
increasing $\lambda$ and for $\lambda=30$ one needs twice as many
iterations until convergence occurs compared to the $\lambda=3$ case.
Interestingly, at a given $\lambda,$ the number of iterations does not
depend on $L.$

\begin{figure}[!t]
\begin{center}
\includegraphics[keepaspectratio,width=0.6\textwidth,angle=0]{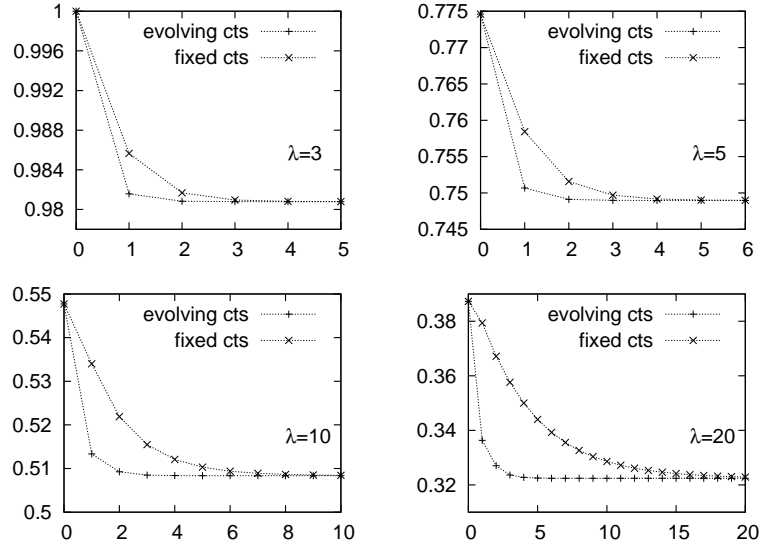}
\caption{A comparison between the convergence of the iterative
  algorithms using fixed and evolving counterterms for a fix value of
  the physical cutoff: $\Lambda/\sqrt{2|m^2|}=200.$ The value
  of $v/\sqrt{2|m^2|}$ is shown as function of the number of iterations for
  different values of the coupling.
\label{fig:comp1}}
\end{center}
\end{figure}

Next, we would like to quantify the extent of achievable improvement
given by the use of the iteratively evolved coupling and mass
counterterms, as compared to the solution of the equations which use
the counterterms originally derived in \cite{patkos08}.  In these two
cases we compare in Fig.~\ref{fig:comp1} the number of iterations
needed for the convergence of $v$ at four different coupling
constants.  We use a reasonably large cutoff
$\Lambda/\sqrt{2|m^2|}=200$, around which the convergent results are
expected to show cutoff independence within some desired accuracy.
One observes the advantage of evolving the counterterms even for
lower values of the renormalized coupling. For larger values
($\lambda\gtrsim 10$) the improvement is significant, as one can see
from the large iteration number needed by the algorithm using fixed
counterterms.  Therefore, it is expected in general that solving
self-consistent equations iteratively in a fully nonperturbative
regime with fixed counterterms is not efficient from a numerical point
of view. For large couplings, $\lambda\gtrsim 10,$ the number of
iterations needed in case of using fixed counterterms are at least a
factor of 2 larger than the number of iterations needed for the
explicitly finite equations to converge.

\begin{figure}[!t]
\begin{center}
\includegraphics[keepaspectratio,width=0.496\textwidth,angle=0]{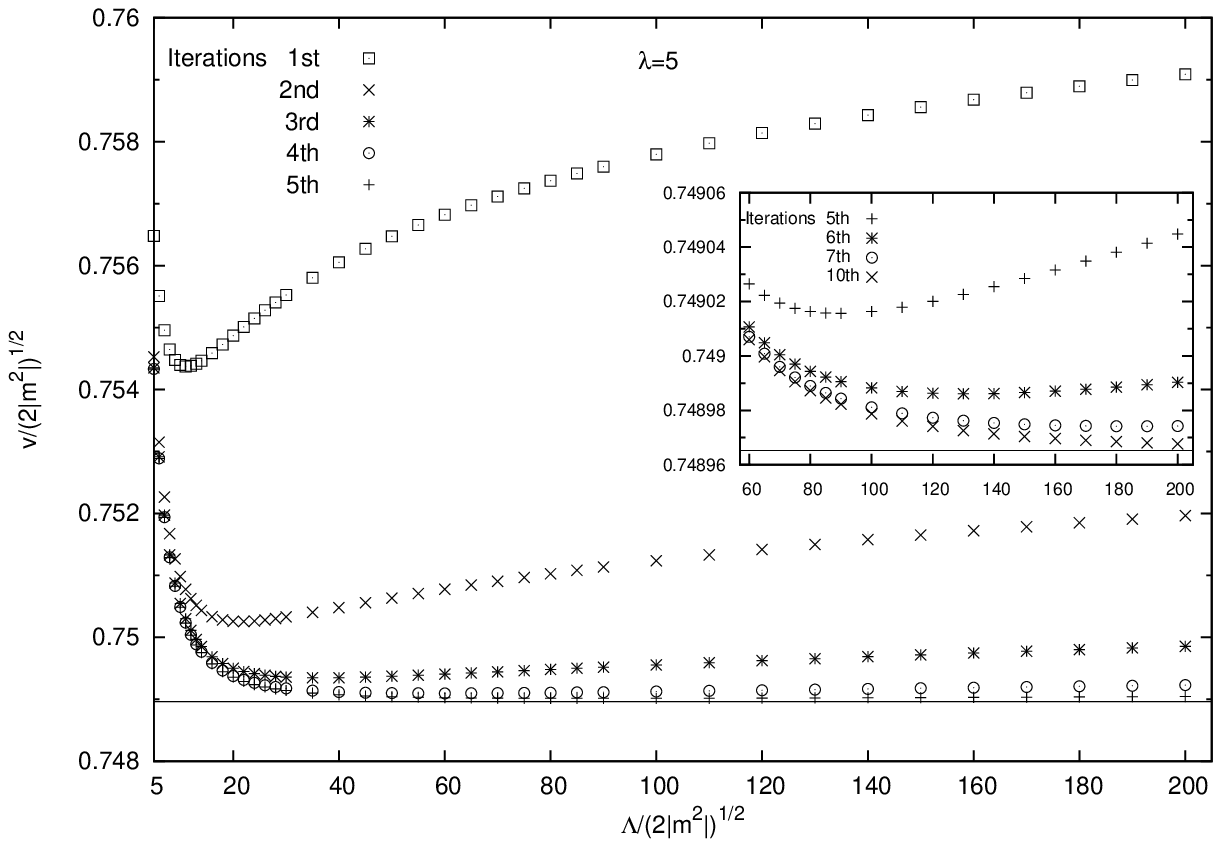}
\includegraphics[keepaspectratio,width=0.496\textwidth,angle=0]{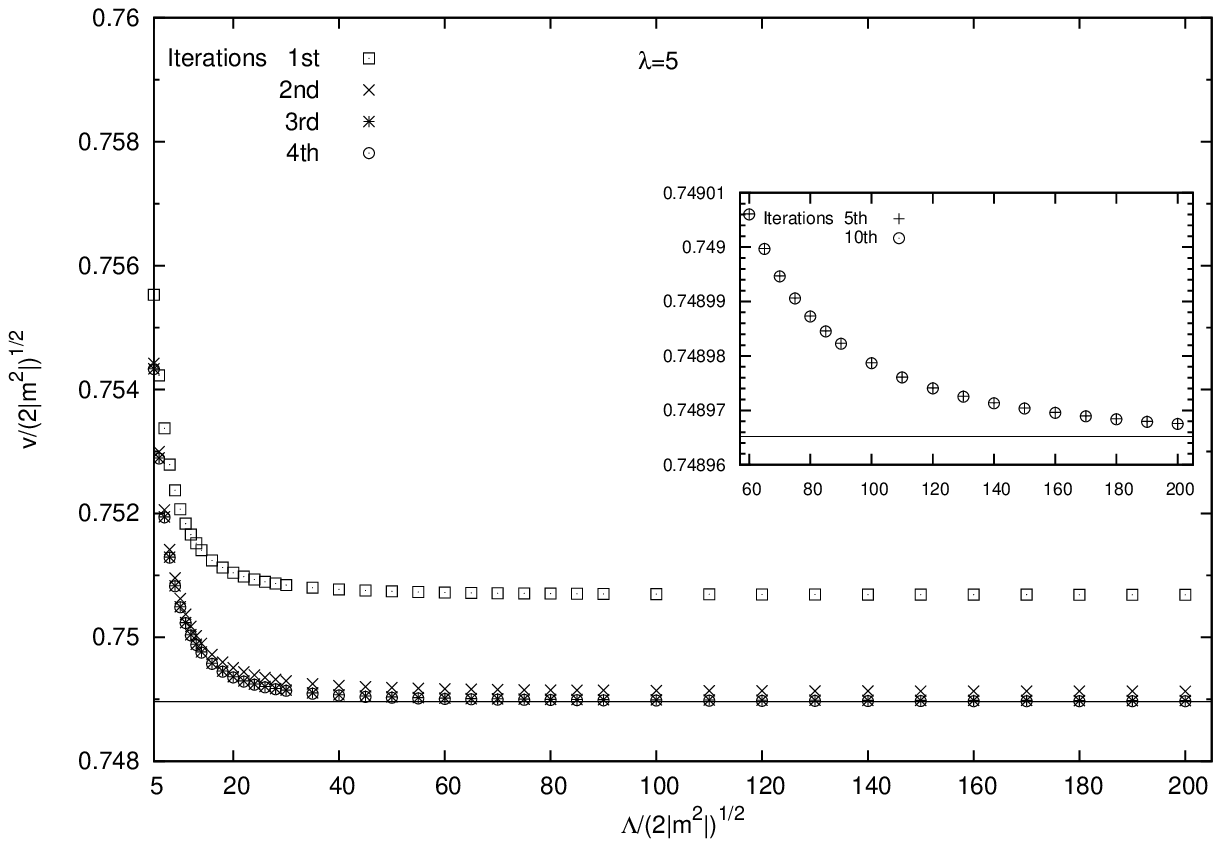}
\caption{
A comparison between the convergence rate of the iterative algorithms
using fixed (left panel) and evolving counterterms (right panel)
as reflected in the change of the vacuum expectation value at
different values of the physical cutoff $\Lambda$. The horizontal line
represents the solution of the explicitly finite system of equations
obtained with $L/\sqrt{2|m^2|}=500,$ where $L$ is the maximal value of
the modulus of the momentum stored on the grid. The insets show the
convergence at large iteration numbers.
\label{fig:comp2}}
\end{center}
\end{figure}

In Fig.~\ref{fig:comp2} we compare for $\lambda=5$ the ways how a
cutoff independent solution is reached using fixed and evolving
counterterms.  The left panel shows that with increasing values of
$\Lambda$ a plateau in the $v(\Lambda)$ functions is approached
rather slowly when the fixed counterterms of \cite{patkos08} are used,
despite the fact that the coupling is not large. Looking at the plot
with values of $\Lambda$ spanning a large range it seams that after
several iterations a plateau is finally reached. However, the
structure in $v(\Lambda),$ better seen in the inset which uses a
smaller $\Lambda$ range, shows that the convergence is not uniform:
more iterations are needed for larger cutoffs to get close to the true
solution, that is the solution of the explicitly finite equations
(horizontal line).  This nonuniformity of the convergence in
$\Lambda$ is more pronounced for larger values of the coupling
constant. Contrary to this, the right panel of Fig.~\ref{fig:comp2}
shows a uniform convergence in $\Lambda$ when the iterative algorithm
uses evolving counterterms.  Since by the very construction of the
evolving counterterms a plateau is observed already for moderate
values of the cutoff at every order of the iteration, one sees again
that with this improved approach the final, convergent result is
obtained in a more efficient way.

\begin{figure}[!t]
\begin{center}
\includegraphics[keepaspectratio,width=0.5\textwidth,angle=0]{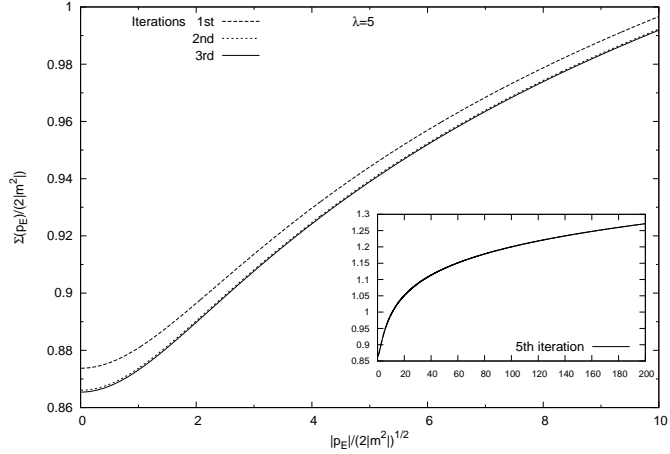}
\caption{
The convergence of the full self-energy obtained for $\lambda=5$ with
evolving counterterms for $\Lambda/\sqrt{2|m^2|}=200.$ The inset
shows the behavior of the self-energy at large momenta.
\label{fig:SE}}
\end{center}
\end{figure}

The convergence of the self-energy can be seen in Fig.~\ref{fig:SE}.
The function converges smoothly, i.e. for all momentum values the
self-energy reaches its converged value almost simultaneously. Note
however that the convergence for $p_E\simeq 0$ is slightly slower as
compared to larger values of the momenta.

\section{Conclusions}

We presented an approach which can be used to obtain a very accurate
numerical solution of the self-consistent propagator equation coupled
to the field equation, both derived from the two-loop level
approximation of the 2PI effective action in the $\phi^4$ model. We
showed that the renormalization method developed in \cite{patkos08}
which removes the divergences using minimal subtraction is equivalent
to imposing nontrivial renormalization conditions on the self-energy,
the curvature of the effective potential and on kernels of
Bethe-Salpeter equations related to these quantities. The use of
renormalization conditions also allowed us to construct explicitly
finite propagator and field equations.

We compared the convergence of the iterative method applied to solve,
on the one hand, the explicitly finite equations and, on the other
hand, the equations which contains the explicitly calculated
counterterms of \cite{patkos08}. The very same results were obtained
numerically which indicates the correctness of our analytic study on
renormalization. It turned out that, especially for larger coupling
constants ($\lambda \gtrsim 10$), the convergence rate of the
algorithm which uses the counterterms is quite poor. This should not
come as a surprise given that the counterterms were determined based
on the asymptotic behavior of the propagator. Within an iterative
procedure the asymptotics develops progressively and is reached only
when the solution converged.  Only then the divergences present in the
equations match the counterterms constructed to cancel them, at
intermediate steps of the iteration process the use of fixed
counterterms results in an oversubtraction.  In order to cure this, we
managed to develop a different algorithm which rederives the
counterterms at every step of the iterative procedure from the
condition that the counterterms cancel the divergent part of the
integrals at every order of the iteration. This method which uses
evolving counterterms turned out to be very effective in obtaining the
converged solution within a few iterative steps also for larger
couplings.

The presented approach and some of the methods applied in our present
work may be directly extended, first to a zero temperature numerical
solution of the model in Minkowski space, and then to a finite
temperature solution of the model, where it would be interesting to
investigate the order of the phase transition.  We expect that the
solution which can be obtained with the method described here could
serve as a good benchmark for other, even more effective methods used
in finite temperature studies (see e.g.
\cite{Berges:2004hn,arrizabalaga06}).  Some of these investigations
will be presented in forthcoming publications.  We expect that our
methods presented here will represent a good basis for obtaining
highly accurate and/or numerically controlled solutions of the 2PI
approximation in more complicated theories as well.  We plan to carry
out the study of the phase transition in the $O(N)$ symmetric model
using large-$N$ expansion. This is motivated by the fact that properly
renormalized complete solutions of the equations derived in the $O(N)$
model at next-to-leading order of the 2PI-$1/N$ expansion are missing
from the literature. The renormalization and the thermodynamics of the
$O(N)$ model was investigated recently at the next-to leading order of
the 1PI-$1/N$ expansion in \cite{Andersen:2008qk}.

\begin{acknowledgments}
The authors thank A. Patk{\'o}s for valuable discussions and his
constant support, and together with A. Jakov\'ac for useful remarks
on the manuscript. They also thank U. Reinosa and J. Serreau
for illuminating discussions on some topics related to this work. This
work is supported by the Hungarian Research Fund (OTKA) under Contract
Nos. K77534 and T068108.
\end{acknowledgments}

\appendix

\section{Integrals appearing in the finite Euclidean equations
\label{app:quantum-corr}}

Since the functions appearing in the quantum correction integrals depend
only on the modulus of the four-momentum, certain parts of the angular
integral and in some cases the entire angular integral can be performed
analytically. The angular integration for the integrals involving $\Dr$
is trivial, while the integral over the modulus of the four-momentum
can be performed only numerically. As explained at the beginning of 
Sec.~\ref{sec:implement} for the convolution of two functions stored
on a grid one uses
\bea
C_L[f,g](k_E)&=&\int \frac{d^4 p_E}{(2\pi)^4}f(|p_E|)g(|p_E+k_E|)
\theta(L-|k_E|) \theta(L-|k_E+p_E|)\nonumber\\
&=&\frac{1}{8\pi^3 k_E^2}
\int_0^L d p\,p\,f(p) \int_{|p-|k_E||}^{\textrm{min}(p+|k_E|,L)} d q\,q\,g(q)
\sqrt{4p^2 q^2-\big (p^2+q^2-k_E^2\big)^2},
\label{Eq:iden1}
\eea
where $\theta$ is the step function, and for the modulus we introduced
$p=|p_E|,$ $q=|k_E+p_E|,$ $L$ is the maximal value of the modulus
stored on the grid. The $q$ integral is a remnant of the angular
integration in a four-dimensional spherical coordinate system and is
obtained with the change of variable
$\cos\theta_1=(q^2-k^2-p^2)/(2 k p).$ When the
function $g$ is $\Dv,$ which is not stored on the grid, that is there
is no need for the second $\theta$ function, the entire angular
integration can be done analytically, and with the change of variable
$t=\tan(\theta_1/2)$ one has
\be
\int_0^\pi d\theta_1 \frac{\sin^2\theta_1}
{p_E^2+k_E^2 + 2 |p_E|\,|k_E|\cos\theta_1+M_0^2}=
\frac{\pi}{4 k_E^2 p_E^2} \left(
k_E^2+p_E^2+M_0^2-\sqrt{\big(k_E^2+p_E^2+M_0^2\big)^2-4 k_E^2 p_E^2}
\right).
\label{Eq:iden2}
\ee 
Using (\ref{Eq:iden2}) in the first term of (\ref{Eq:sigmar-eucl}) 
and for the setting-sun contribution to the field equation 
(\ref{Eq:EoS-fin-eucl}) which contains one vacuum propagator one obtains
\bea
\label{Eq:I_mv}
I^\textrm{(mv)}(k_E)&=&
-\frac{1}{16\pi^2 k_E^2}\int_0^L d p\,p \left(
k_E^2+p^2+M_0^2- \sqrt{\big(k_E^2+p^2+M_0^2\big)^2-4k_E^2p^2}
\right)\Dm(p),\\
S^\textrm{(mmv)}&=&
-\frac{1}{128\pi^4}\int_0^L d k\,k\,\Dm(k)
\int_0^L d p\,p 
\left(k^2+p^2+M_0^2-\sqrt{\big(k^2+p^2+M_0^2\big)^2-4 k^2 p^2}\right)
\Dm(p).\ \ 
\eea
For the second term of (\ref{Eq:sigmar-eucl}) and the setting-sun 
contribution to (\ref{Eq:EoS-fin-eucl}) which contains only matter 
propagators one uses (\ref{Eq:iden1}) and the results of the angular
integrations are
\bea
I^\textrm{(mm)}(k_E)&=&
-\frac{1}{8\pi^3 k_E^2}\int_0^L d p\,p\,\Dm(p)
\int_{|p-|k_E||}^{\textrm{min}(p+|k_E|,L)} 
d q\,q\,\sqrt{4 p^2 q^2-\big(p^2+q^2-k_E^2\big)^2} \Dm(q),\\
S^\textrm{(mmm)}&=&-\frac{1}{64\pi^5}\int_0^L d k\,k\,\Dm(k)\int_0^L 
d p\,p\,\Dm(p)\int_{|p-k|}^{\textrm{min}(p+k,L)} d q\,q 
\sqrt{4 p^2q^2-(p^2+q^2-k^2)^2}
\Dm(q).\ \ \ 
\eea

\section{Cutoff dependence of divergent integrals \label{app:cutoff}}

For reader's convenience we summarize below the equations for the
counterterms obtained in \cite{patkos08} and rederived in
Sec.~\ref{sec:finite_equation} by imposing appropriate
renormalization conditions:
\begin{subequations}
\bea
0=&&\delta m_2^2+\frac{1}{2}(\lambda+\delta\lambda_0)
\left[T_d^{(2)}+(m^2-M_0^2)
T_d^{(0)}\right],
\label{Eq:delta-m22}
\\
0=&&\delta\lambda_0+\frac{1}{2}\lambda(\lambda+\delta\lambda_0)T_d^{(0)},
\label{Eq:delta-lambda-0}
\\
0=&&\delta\lambda_2+\frac{1}{2}\lambda(\lambda+\delta\lambda_0)
\left(T_d^{(0)}+\lambda T_d^{(I)}\right)+\lambda^2T_d^{(0)},
\label{Eq:delta-lambda-2}
\\
0=&&\delta m^2_0+\frac{1}{2}(\lambda+\delta\lambda_2)
\left[T_d^{(2)}+(m^2-M_0^2) T_d^{(0)}\right]+\frac{1}{2}\lambda^2(m^2-M_0^2)
\left((T_d^{(0)})^2+T_d^{(I)}\right)+
\frac{1}{6}\lambda^2S_{0}(0),
\label{Eq:delta-m02}
\\
0=&&\delta\lambda_4+\frac{3}{2}\lambda(\lambda+\delta\lambda_2+\lambda^2
T_d^{(0)})\left(T_d^{(0)}+\lambda T_d^{(I)}\right)+\frac{3}{2}\lambda^3 
\left(T_d^{(I)}+\lambda T_d^{(I,2)}\right).
\label{Eq:delta-lambda-4}
\eea
\label{Eq:all_counterterms}
\end{subequations}
\vspace{-2mm}   

\noindent
The first three and the last two expressions come from the
renormalization of the propagator equation and field equation,
respectively. The divergent quantities $T_d^{(2)},$ $T_d^{(0)},$ and
$S_{0}(0)$ are the vacuum parts of the tadpole integral
(\ref{Eq:T_def}), the bubble integral (\ref{Eq:I_def}) at vanishing
external momentum and the setting-sun integral (\ref{Eq:SS_def}), all
calculated with the propagator $G_{0}$ given in (\ref{Eq:G0}):
\begin{subequations}
\bea
T_d^{(2)}&=&\int_P G_{0}(P),\\
T_d^{(0)}&=&\frac{d T_d^{(2)}}{d M_0^2} =-i\int_P G_{0}^2(P),\\
S_{0}(0)&=&-i\int_K \int_Q G_{0}(K) G_{0}(Q)G_{0}(K+Q).
\eea 
\label{Eq:Td2_Td0_S0}
\end{subequations}
\vspace{-2mm}   

\noindent
The remaining two divergent integrals $T_d^{(I)}$ and $T_d^{(I,2)}$
are defined in terms of the finite part of the bubble 
integral (\ref{Eq:I_def}) calculated with the propagator $G_{0},$ that is
\be
I_{0,F}(K)=-i\int_P \Big(G_0(P) G_0(P+K)-G_0^2(P)\Big).
\label{Eq:fin-bubble}
\ee
and read
\begin{subequations}
\bea
T_d^{(I)}&=&-i\int_K G_0^2(K) I_{0,F}(K)=
-i\int_K G_0^2(K) I_0(K) - \big(T_d^{(0)}\big)^2
,\\ 
T_d^{(I,2)}&=&-i\int_K G_0^2(K) I_{0,F}^2(K)=
-i\int_K G^2_0(K) I^2_0(K) - 2 T_d^{(I)} T_d^{(0)} - \big(T_d^{(0)}\big)^3.
\eea
\label{Eq:T_dI_T_dI2}
\end{subequations}
\vspace{-2mm}   

\noindent
With a four-dimensional rotational invariant cutoff regularization in Euclidean
space, the divergent integrals are computed as follows
\begin{subequations}
\bea
T_d^{(2)}&=&\frac{\Lambda^2}{16\pi^2}\left[1-
\frac{M_0^2}{\Lambda^2}\log\left(1+\frac{\Lambda^2}{M_0^2}\right)\right],
\quad
T_d^{(0)}=\frac{1}{16\pi^2}\left[
\frac{\Lambda^2}{\Lambda^2+M_0^2}-\log\left(1+\frac{\Lambda^2}{M_0^2}\right)
\right],\\
S_0(0)&=&-\frac{1}{8\pi^2}\int_0^\Lambda d k\frac{k^3}{(k^2+M_0^2)} C_\Lambda[\Dv,\Dv](k),\\
T_d^{(I)}&=&\frac{1}{8\pi^2}\int_0^\Lambda d k\frac{k^3}{(k^2+M^2_0)^2}
C_\Lambda[\Dv,\Dv](k)-\big(T_d^{(0)}\big)^2,
\\
T_d^{(I,2)}&=&-\frac{1}{8\pi^2} \int_0^\Lambda d k\frac{k^3}{(k^2+M^2_0)^2}
\Big(C_\Lambda[\Dv,\Dv](k)\Big)^2 - 2 T_d^{(I)} T_d^{(0)} 
- \big(T_d^{(0)}\big)^3,
\eea
\label{Eq:div_ints_euclid}
\end{subequations}
\vspace{-2mm}   

\noindent
where for $C_\Lambda[\Dv,\Dv](k)$ one uses (\ref{Eq:iden1})
with $\Dv(k)=1/(k^2+M_0^2)$ and with $L$ replaced by the physical
cutoff $\Lambda.$ The counterterms determined from
(\ref{Eq:all_counterterms}) using the expressions in
(\ref{Eq:div_ints_euclid}) cancel the divergences of tadpole, bubble
and setting-sun integrals, which in Euclidean space read as
\be
T[\Delta]=\frac{1}{8\pi^2}\int_0^\Lambda d k\, k^3 \Delta(k),\quad
I(k,\Delta)=-C_\Lambda[\Delta,\Delta](k),\quad
S(0)=-\frac{1}{8\pi^2}\int_0^\Lambda d k\,k^3
\Delta(k) C_\Lambda[\Delta,\Delta](k).
\ee

\end{document}